\documentclass[a4paper,12pt,dvipsnames,usernames]{article}
\pdfoutput=1

\usepackage{jheppub} 

\usepackage[utf8]{inputenc}
\usepackage{graphicx}
\usepackage{amsmath}
\usepackage{amssymb}
\usepackage{bm}
\usepackage{paralist,array}
\usepackage{units} 
\graphicspath{{fig/}{./Plots/}} 
\usepackage{slashed}
\usepackage{todonotes}
\usepackage[parfill]{parskip}
\usepackage{longtable}
\usepackage{subfig}

\title{Phenomenology of GeV-scale scalar portal}

\author[1]{Iryna~Boiarska,} 
\author[2]{Kyrylo~Bondarenko,} 
\author[2]{Alexey~Boyarsky,} 
\author[3]{Volodymyr~Gorkavenko,}
\author[2]{Maksym~Ovchynnikov,}
\author[4]{Anastasia~Sokolenko}

\affiliation[1]{Discovery Center, Niels Bohr
   Institute, Copenhagen University, Blegdamsvej 17, DK-2100, Copenhagen,  Denmark}
\affiliation[2]{Intituut-Lorentz, Leiden University, Niels Bohrweg 2, 2333 CA, Leiden, The Netherlands}
\affiliation[3]{Department of Physics, Taras Shevchenko National University of Kyiv, 64 Volodymyrs’ka str. 01601, Kyiv, Ukraine}
\affiliation[4]{Department of Physics, University of Oslo, Box 1048, NO-0371, Oslo, Norway}

\emailAdd{boiarska@nbi.ku.dk}
\emailAdd{bondarenko@lorentz.leidenuniv.nl}
\emailAdd{boyarsky@lorentz.leidenuniv.nl}
\emailAdd{gorkavol@gmail.com}
\emailAdd{ovchynnikov@lorentz.leidenuniv.nl}
\emailAdd{anastasia.sokolenko@fys.uio.no}
  
\begin{document}

\abstract{We review and revise the phenomenology of the scalar portal -- a new scalar particle with the mass in GeV range that mixes with the Higgs boson. In particular, we consider production channels $B\to S K_1(1270)$ and $B\to S K_0^*(700)$ and show that their contribution is significant. We extend the previous analysis by comparing the production of scalars from decays of mesons, of the Higgs bosons and direct production via proton bremsstrahlung, deep inelastic scattering and coherent scattering on nuclei. Relative efficiency of the production channels depends on the energy of the beam and we consider the energies of DUNE, SHiP and LHC-based experiments. We present our results in the form directly suitable for calculations of experimental sensitivities.}

\maketitle

\newpage

\section{Introduction}
 We review and revise the phenomenology of the scalar portal -- a gauge singlet scalar particle $S$ that couples to the Higgs boson and can play a role of a mediator between the Standard model and a dark sector (see e.g.~\cite{Bird:2004ts,Pospelov:2007mp,Krnjaic:2015mbs}) or be involved in the cosmological inflation~\cite{Shaposhnikov:2006xi,Bezrukov:2009yw,Bezrukov:2013fca}. We focus here on the mass range $\lesssim 10\text{ GeV}$ (see however section~\ref{sec:quartic_coupling} for a discussion of larger masses). 

\begin{figure}[h!]
    \centering
    \subfloat[]{\includegraphics[width=0.48\textwidth]{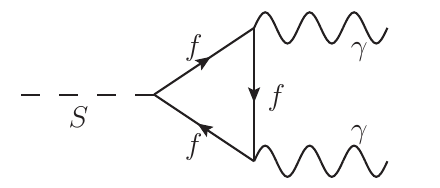}}~\subfloat[]{\includegraphics[width=0.48\textwidth]{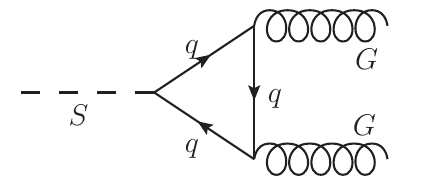}} \\
    \subfloat[]{\includegraphics[height=0.3\textwidth]{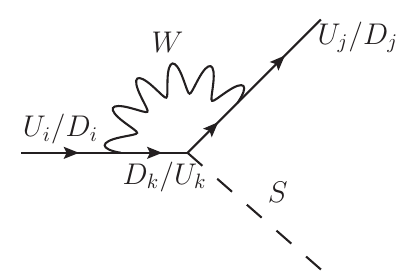}}
    \caption{Examples of effective interactions of the scalar with photons (a), gluons (b), and flavor changing quark operators (c). (See Appendix~\ref{sec:effective-interactions} for details.)}
    \label{fig:Seffective}
\end{figure}

\begin{figure}[h!]
    \centering
    \includegraphics[height=0.42\textwidth]{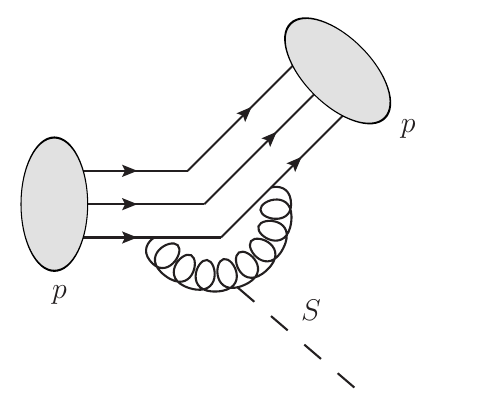}
    \caption{An example of a diagram for the effective interaction of a proton with a scalar, see Appendix~\ref{sec:effective-interactions-nucleons} for details.}
    \label{fig:Sproton}
\end{figure}

The interaction of the $S$ particle with the Standard model particles is similar to the interaction of a light Higgs boson but is suppressed by a small mixing angle $\theta$.  
Namely, the Lagrangian of the scalar portal is
\begin{equation}
\mathcal{L} = \mathcal{L}_{SM}  + \frac{1}{2} \partial_{\mu} S \partial^{\mu} S + 
(\alpha_1 S + \alpha_{2} S^2) (H^{\dagger} H) -\frac{m_{S}^{2}}{2}S^2.
\label{eq:L1}
\end{equation}
 After the electroweak symmetry breaking the Higgs doublet gains a non-zero vacuum expectation value $v$. As a result, the $SHH$ interaction~\eqref{eq:L1} provides a mass mixing between  $S$ and the Higgs boson $h$. 
  Transforming the Higgs field into the  mass basis, $h \to h + \theta S$, one arrives at the following interaction of $S$ with the SM fermions and gauge bosons:
\begin{align}
& \mathcal{L}_{SM}^{S} = -\theta\frac{m_f}{v} S \bar{f}f +
2\theta\frac{ m_W^2}{v} S W^+ W^- +
\theta\frac{m_Z^2}{v} S Z^2 + \alpha\left(\frac{1}{4v}S^2 h^2 + \frac{1}{2}S^2 h\right),
\label{g01}
\end{align}
where $\alpha \equiv 2\alpha_{2} v$. 
These interactions also mediate effective couplings of the scalar to photons, gluons, and flavor changing quark operators, see Fig.~\ref{fig:Seffective}. 
Additionally, the effective proton-scalar interaction that originates from the interaction of scalars with quarks and gluons (see Fig.~\ref{fig:Sproton}) will also be relevant for our analysis.
The effective Lagrangian for these interactions is discussed in Appendix~\ref{sec:effective-interactions}.

Searches for light scalars have been previously performed by CHARM, LHCb and Belle~\cite{Schmidt-Hoberg:2013hba,Clarke:2013aya}, CMS~\cite{Sirunyan:2018owy} and ATLAS~\cite{Aad:2015txa,Aaboud:2018sfi} experiments. Significant progress in searching for light scalars can be achieved by the proposed and planned intensity-frontier experiments SHiP~\cite{Alekhin:2015byh,Anelli:2015pba,SHiP:2018xqw}, CODEX-b~\cite{Gligorov:2017nwh}, MATHUSLA~\cite{Curtin:2018mvb,Chou:2016lxi,Curtin:2017izq,Evans:2017lvd,Helo:2018qej,Lubatti:2019vkf}, FASER~\cite{Feng:2017uoz,Feng:2017vli,Kling:2018wct}, SeaQuest~\cite{Berlin:2018pwi},  NA62~\cite{Mermod:2017ceo,CortinaGil:2017mqf,Drewes:2018gkc}, DUNE~\cite{Adams:2013qkq}. 

The phenomenology of light GeV-like scalars has been studied in~\cite{Bezrukov:2009yw,Clarke:2013aya,Bezrukov:2018yvd,Batell:2009jf,Winkler:2018qyg,Monin:2018lee,Bird:2004ts}, and in~\cite{Voloshin:1985tc,Raby:1988qf,Truong:1989my,Donoghue:1990xh,Willey:1982ti,Willey:1986mj,Grzadkowski:1983yp,Leutwyler:1989xj,Haber:1987ua,Chivukula:1988gp} in the context of a light Higgs boson. However, in the literature, there are still conflicting results, both for the scalar production and decay. In this work, we reanalyze the phenomenology of light scalars and present the results in the form directly suitable for experimental sensitivity estimates.

\section{Scalar production}
\label{sec:production}

\subsection{Mixing with the Higgs boson}

In this section, we will discuss the scalar production channels that are defined by the mixing between a scalar and the Higgs boson.

\begin{figure}[t!]
    \centering
    \subfloat[]{\includegraphics[height=0.25\textwidth]{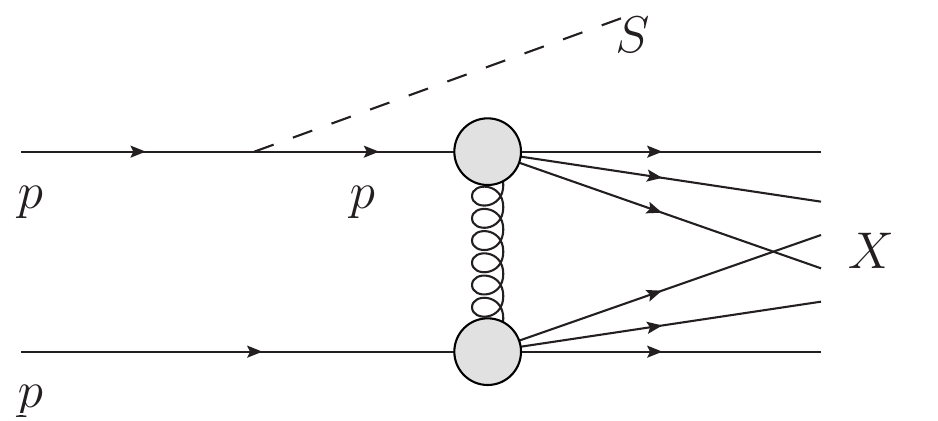}}
    \\
    \subfloat[]{\raisebox{-0.7\height}{\includegraphics[width=0.36\textwidth]{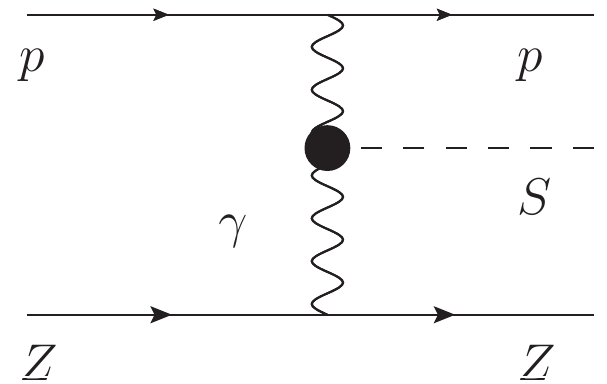}}~\raisebox{-0.5\height}{\includegraphics[width=0.62\textwidth,trim={0 0 12cm 0},clip]{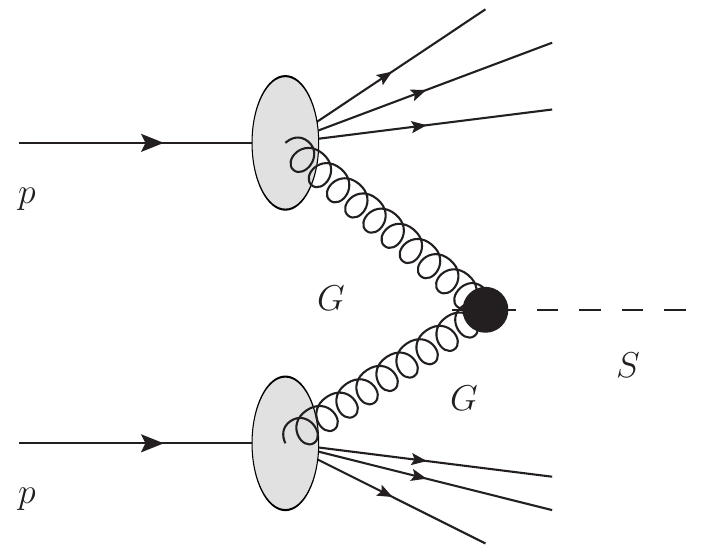}}} \\
    \subfloat[]{\includegraphics[height=0.38\textwidth]{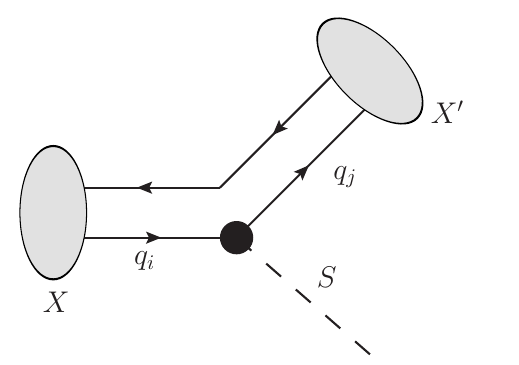}}
    \caption{Example diagrams for the main production channels of a scalar $S$ production in proton-nucleus collisions: proton bremsstrahlung (a), photon and gluon fusion (b), decay of secondary mesons (c).}
    \label{fig:Sproduction}
\end{figure}

In proton-proton or proton-nucleus collisions, a scalar particle: (a) can be emitted by the proton, (b) produced from photon-photon, gluon-gluon or quark-antiquark fusion in proton-proton or proton-nucleus interactions or (c) produced in the decay of the secondary particles, see Fig.~\ref{fig:Sproduction}.
Let us compare these three types of the scalar production mechanisms depending on the collision energy and the scalar mass. In the following we will present the results for three referent proton-proton center-of-mass energies: $\sqrt{s_{pp}} \approx 16\text{ GeV}$ (corresponding to the beam energy of the DUNE experiment), $\sqrt{s_{pp}} \approx 28\text{ GeV}$ (SHiP) and $\sqrt{s_{pp}} = 13\text{ TeV}$ (LHC).

\textbf{The proton bremsstrahlung (the case (a))} is a process of a scalar emission by a proton in proton-proton interaction. For small masses of  scalars, $m_S < 1$~GeV, this is a usual bremsstrahlung process  described by  elastic nucleon-scalar interaction with a coupling constant $\theta g_{SNN}\sim \theta m_{p}/v$, see Appendix~\ref{sec:effective-interactions-nucleons}. However, the probability of elastic interaction decreases with the scalar mass and we need to take into account inelastic processes. The probability for the bremsstrahlung is
\begin{equation}
    P_{\text{brem}} = \theta^{2} g_{SNN}^{2} \mathcal{P}_{\text{brem}}(m_{S},s_{pp}),
    \label{eq:Pbrem}
\end{equation}
where $\mathcal{P}_{\text{brem}}$ is a proton bremsstrahlung probability for the case $\theta=g_{SNN}=1$ (see Appendix~\ref{sec:bremsstrahlung}). This quantity varies from $10^{-2}$ for DUNE and SHiP to $10^{-1}$ for the LHC, see Appendix~\ref{sec:bremsstrahlung}. In this estimate, we neglect possible effects of QCD scalar resonances such as $f_{0}(500), f_{0}(580), f_{0}(1370),\dots$, that could resonantly enhance the scalar production for some masses (see~\cite{Batell:2020vqn,Foroughi-Abari:2021zbm} and Appendix~\ref{sec:effective-interactions-nucleons},~\ref{sec:bremsstrahlung} for further details).

\begin{table}[t]
\centering
\begin{tabular}{|c|c|c|c|}
 \hline  
 Experiment  & DUNE & SHiP & LHC \\
 & $(\sqrt{s_{pp}} = 16\text{ GeV})$ & $(\sqrt{s_{pp}} = 28\text{ GeV})$ & $(\sqrt{s_{pp}} = 13\text{ TeV})$\\
 \hline
 $W_{GG}(m_S=2\text{ GeV})$ & $1.2\cdot 10^{3}$ & $1.4\cdot 10^{3}$  & $6.2\cdot 10^{10}$ \\
 \hline
 $W_{GG}(m_S=5\text{ GeV})$ & $1.5\cdot 10^{-1}$ & $7.9$ & $3.9 \cdot 10^{10}$\\
 \hline
\end{tabular}
    \caption{Factors $W_{GG}$ (see Eq.~\eqref{eq:DISsigma}) for the DUNE, SHiP and LHC experiments.}
    \label{tab:WDIS}
\end{table}

\begin{figure}[h!]
    \centering
    \includegraphics[width=0.6\textwidth]{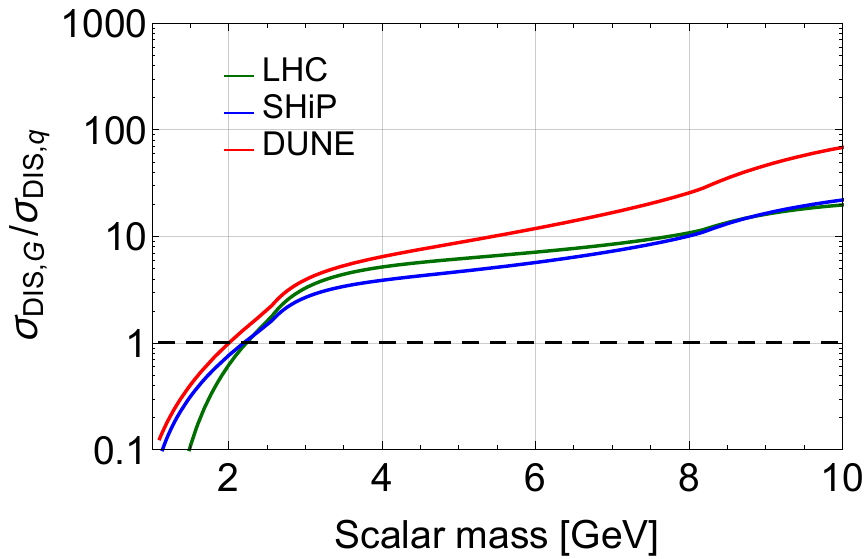}
    \caption{The ratio of cross sections of the scalar production in deep inelastic scattering via gluon and quark fusion. The dashed line corresponds to a ratio equal to unity. ``LHC'', ``SHiP'' and ``DUNE'' denote correspondingly the results for the proton-proton center-of-mass energies $\sqrt{s_{pp}} = 13\text{ TeV}$, $\sqrt{s_{pp}} = 28\text{ GeV}$ and $\sqrt{s_{pp}} = 16\text{ GeV}$.}
    \label{fig:dis-production-plot}
\end{figure}

\textbf{For the case (b)}, we have to distinguish the photon-photon fusion that can occur for an arbitrary transferred momentum and, therefore, an arbitrary scalar mass (as electromagnetic interaction is long-range), and gluons or quark-antiquarks fusion (the so-called deep inelastic scattering processes (DIS)), which is effective only for $m_S \gtrsim 1 \text{ GeV}$.
The scalar production in the DIS process can be estimated as $P_{\text{DIS}} = (\sigma_{\text{DIS},G}+\sigma_{\text{DIS},q})/\sigma_{pp}$, where $\sigma_{pp}$ is the total proton-proton cross section and $\sigma_{\text{DIS},X}$ is the cross section of scalar production in the DIS process,
\begin{equation}
    \sigma_{\text{DIS},G} \sim \frac{\theta^2 \alpha_{s}^2(m_S) m_S^2}{s_{pp} v^2} W_{GG}(m_S,s_{pp}),
    \quad
    \sigma_{\text{DIS},q} \sim \frac{\theta^2 m_q^2}{s_{pp} v^2} W_{q\bar{q}}(m_S,s_{pp}).
    \label{eq:DISsigma}
\end{equation}
Here, $\sqrt{s_{pp}}$ is the center-of-mass energy of colliding protons and $W_{XX}$ given by Eq.~\eqref{eq:partonic-weight} is a dimensionless combinatorial factor that, roughly, counts the number of the parton pairs in  two protons that can make a scalar. The values of $W_{XX}$ factors for some scalar masses and experimental energies are given in Table~\ref{tab:WDIS}. In Fig.~\ref{fig:dis-production-plot} we show the ratio between cross sections of gluon-gluon and quark-antiquark fusion. We see that quark fusion is relevant only for low scalar masses, while for $m_S \gtrsim 2$~GeV the gluon fusion dominates for all collision energies considered.

In the case of \emph{the production of a scalar in photon fusion}, the most interesting process is the coherent scattering on the whole nucleus, as its cross section is enhanced by a factor $Z^2$, where $Z$ is the charge of the nucleus. 
The electromagnetic process $p+Z \to p+Z+S$ involves the effective $S\gamma\gamma$ vertex proportional to $\theta \alpha_{\text{EM}}$, see Appendix~\ref{sec:effective-interactions-gauge-bosons}. The probability of the process is $P_{\gamma\text{ fus}} = \sigma_{\gamma\text{ fus}}/\sigma_{pZ}$, where the fusion cross section $\sigma_{\gamma\text{ fus}}$ has a structure similar to that of gluon fusion~\eqref{eq:DISsigma}:
\begin{equation}
    \sigma_{\gamma\text{ fus}} \sim 10^{-2} \frac{\theta^2 Z^{2}\alpha_{\text{EM}}^{4}m_{S}^{2}}{v^{2}s_{pZ}} W_{\text{coh}},
\end{equation}
where $\sqrt{s_{pZ}}$ is the CM energy of the proton and nucleus, and $W_{\text{coh}}$ given by Eq.~\eqref{eq:coherent-weight} is a dimensionless combinatorial factor that counts the number of pair of photons that can form a scalar. It ranges from $10^{6}$ for the DUNE energies to $10^{14}$ for the LHC energies.

Let us compare the probabilities of  photon fusion and  proton bremsstrahlung,
\begin{align}
    \frac{P_{\gamma\text{ fus}}}{P_{\text{brem}}} &\sim 10^{-2}\frac{Z^{2}}{s_{pZ}\sigma_{pZ}} 
    \frac{\alpha_{\text{EM}}^{4}}{g_{SNN}^{2}} 
    \frac{m_{S}^{2}}{v^{2}}
    \frac{W_{\text{coh}}}{\mathcal{P}_{\text{brem}}} 
    \sim \\ &\sim 
    10^{-15}\frac{(100\text{ GeV})^{2}}{s_{pp}}\frac{Z^{2}}{A^{1.77}}\left(\frac{m_{S}}{1\text{ GeV}}\right)^{2}\frac{W_{\text{coh}}}{\mathcal{P}_{\text{brem}}} \lesssim 10^{-4}
\end{align}
for all three energies considered. Here we used $s_{pZ}\approx A s_{pp}$, where $A$ is the nucleus mass number. The proton-nucleus cross section $\sigma_{pZ}$ weakly depends on energy and can be estimated  as $\sigma_{pZ} \simeq 50\text{ mb}\  A^{0.77}$~\cite{Tanabashi:2018oca,Carvalho:2003pza}. This ratio is smaller than $10^{-4}$ for all energies and scalar masses of interest. Next, comparing the probabilities of the production in photon fusion and in DIS, we obtain
\begin{equation}
    \frac{P_{\gamma 
\text{ fus}}}{P_{\text{DIS}}}\sim \frac{Z^2 \alpha_{\text{EM}}^4}{\alpha_s^2} \frac{s_{pp}}{s_{pZ}} \frac{\sigma_{pp}}{\sigma_{pZ}} \frac{W_{\text{coh}}}{W_{\text{DIS}}} \sim 10^{-8}\frac{Z^{2}}{A^{1.77}}\frac{W_{\text{coh}}}{W_{\text{DIS}}} \lesssim 10^{-4},
\end{equation}
where we used that $s_{pZ}/s_{pp}\approx A$ and $W_{\text{coh}}/W_{\text{DIS}} \lesssim 1$ for all three energies considered, see Appendix~\ref{sec:production-direct-coherent}. The proton-proton cross section also  depends on the energy  weakly,
and we can estimate   $\sigma_{pZ}/\sigma_{pp}\sim A^{0.77}$
(see Appendix~\ref{sec:bremsstrahlung}).

\emph{We conclude that the scalar production in photon fusion is always sub-dominant for the considered mass range of scalar masses and beam energies.}

Let us now compare gluon fusion and proton bremsstrahlung with the \textbf{production from secondary mesons (type (c))}.
The latter can be roughly estimated using ``inclusive production'', i.e. production from the decay of a free heavy quark, without taking into account that in reality this quark is a part of different mesons with different masses.  
This is only an order of magnitude estimate that breaks down for $m_S \gtrsim m_q -\Lambda_{\text{QCD}}$, so it can be used only for $D$ and $B$ mesons. We will see however that such an estimate is sufficient to conclude that for the energies of SHiP and LHC the production from mesons dominates and we need to study it in more details (see more detailed analysis below).

The inclusive branching $\text{BR}_{\text{incl}}(X_{Q_{i}} \to X_{Q_{j}}S)$ can be estimated using the corresponding quark level process $Q_{i} \to Q_{j} S$. To minimize QCD uncertainty we follow~\cite{Chivukula:1988gp,Curtin:2018mvb} and estimate the inclusive branching ratio as
\begin{equation}
    \text{BR}_{\text{incl}}(X_{Q_{i}} \to X_{Q_{j}}S) \simeq \frac{\Gamma(Q_{i}\to Q_{j}S )}{\Gamma(Q_{i} \to Q' e\bar{\nu}_{e})}\times\text{BR}_{\text{incl}}(X\to X_{Q'}e\bar{\nu}_{e}),
    \label{eq:prodBRinclusiveB}
\end{equation}
where $\Gamma(Q_{i}\to Q'e\bar{\nu}_{e})$ is the semileptonic decay width of a quark $Q_{i}$ calculated using the Fermi theory and $\text{BR}_{\text{incl}}(X\to X_{Q'}e\bar{\nu}_{e})$ is the experimentally measured inclusive branching ratio. As both the quark decay widths in~\eqref{eq:prodBRinclusiveB} get the QCD corrections, their total effect in~\eqref{eq:prodBRinclusiveB} is expected to be minimal~\cite{Chivukula:1988gp}

For $D$ and $B$ mesons decays the inclusive production probabilities are~\cite{Chivukula:1988gp,Evans:2017lvd}
\begin{align}
    P_{D} &\sim 2\chi_{c\bar{c}} \times \text{BR}(c\to Su) \sim 6\cdot 10^{-11} \ \theta^{2} \chi_{c\bar{c}} \left( 1-\frac{m_{S}^{2}}{m_{c}^{2}}\right)^{2},
    \label{eq:PD}
    \\
    P_{B} &\sim 2\chi_{b\bar{b}} \times \text{BR}(b\to Ss) \sim 13 \ \theta^{2}\chi_{b\bar{b}}\left( 1-\frac{m_{S}^{2}}{m_{b}^{2}}\right)^{2},
    \label{eq:PB}
\end{align}
where $\chi_{q\bar{q}}$ is the production fraction of the $q\bar{q}$ pair in $pp$ collisions, see Table~\ref{tab:meson-amounts}. The difference in $10^{-11}$ orders of magnitude is mostly coming from  $(m_b/m_t)^4\sim 10^{-7}$  and $|V_{ub}|^2/|V_{ts}|^2\sim 10^{-2}$ (see Appendix~\ref{sec:fcnc} for details). 
In fact, for $D$ mesons the leptonic decay $D\to S e\nu$ with $\text{BR}(D\to S e\nu)\sim 5\cdot 10^{-9}\theta^{2}$ is more important than~\eqref{eq:PD}, see Appendix~\ref{sec:leptonic-decays} for details. We see that the production from $D$ mesons may be important only if the number of $B$ mesons is suppressed by $10^9$ times, which is possible only if the center-of-mass energy of $p$-$p$ collisions is close to the $b\bar{b}$ pair production threshold. 

Let us compare the production from $B$ mesons with the production from proton bremsstrahlung and DIS. Using Eqs.~\eqref{eq:Pbrem},~\eqref{eq:DISsigma} and~\eqref{eq:PB} for masses of scalar below $b$ quark kinematic threshold we get
\begin{equation}
  \frac{P_{\text{brem}}}{P_{B}} \sim \frac{g_{SNN}^2}{\text{BR}(b\to Ss)} \frac{\mathcal{P}_{\text{brem}}}{\chi_{b\bar{b}}} \sim 10^{-7} \frac{\mathcal{P}_{\text{brem}}}{\chi_{b\bar{b}}},
  \label{eq:Pbrem2B}
\end{equation}
\begin{equation}
    \frac{P_{\text{DIS}}}{P_{B}} 
    \sim 10^{-6} \frac{1}{s_{pp} \sigma_{pp}} \left(\frac{m_S}{1\text{ GeV}}\right)^2
    \frac{W_{GG}(m_S,s_{pp})}{\chi_{b\bar{b}}}.
    \label{eq:pDIS-to-B}
\end{equation}

The ratios~\eqref{eq:pDIS-to-B} and ~\eqref{eq:Pbrem2B} depend on the center-of-mass energy of the experiment
(see  Tables~\ref{tab:WDIS} and~\ref{tab:meson-amounts}).

\emph{We conclude that for the experiments with high beam energies, like SHiP or LHC, the most relevant production channel is a production of scalars from secondary mesons. For experiments with smaller energies like, e.g., DUNE the dominant channel is the direct production  of scalars in proton bremsstrahlung and in DIS.}

\begin{table}[t]
\centering
\begin{tabular}{|c|c|c|c|}
 \hline  
 Experiment  & DUNE & SHiP & LHC \\
 & $(\sqrt{s_{pp}} = 16\text{ GeV})$ & $(\sqrt{s_{pp}} = 28\text{ GeV})$ & $(\sqrt{s_{pp}} = 13\text{ TeV})$\\
 \hline
 $\chi_{c\bar{c}}$ & $1.0\cdot 10^{-4}$ & $3.9\cdot 10^{-3}$  & $2.9\cdot 10^{-2}$ \\
 \hline
 $\chi_{b\bar{b}}$ & $1.0\cdot 10^{-10}$ & $2.7\cdot 10^{-7}$ & $8.6 \cdot 10^{-3}$\\
 \hline
\end{tabular}
    \caption{Production fractions of the $q\bar{q}$ pair, $\chi_{q\bar{q}} = \sigma_{q\bar{q}}/\sigma_{pp}$, for the DUNE, SHiP and LHC experiments. We took the production fractions for the DUNE and SHiP experiments from~\cite{Lourenco:2006vw,CERN-SHiP-NOTE-2015-009}. To estimate the production fractions for the LHC, we calculated the total cross section of $B$ and $D$ production using FONLL~\cite{Cacciari:1998it} and took the total cross section of the $pp$ collisions at the LHC energies from~\cite{Tanabashi:2018oca}.}
    \label{tab:meson-amounts}
\end{table}

\textbf{Production from decays of different mesons.}\\
Let us discuss the production of scalars from decays of mesons in more details.
The calculation of branching ratios for two-body decays of mesons is summarized in  Appendix~\ref{sec:meson-two-body-decays}. Above we made an estimate for the cases of $D$ and $B$ mesons that are the most efficient production channels for larger masses of $S$. 
Instead, for scalar masses $m_S<m_K-m_\pi$ the main production channel is the decay of kaons, $K \to S \pi$, see Table~\ref{tab:BR} with the relevant information about these production channels. Numerically, the branching ratio of the production from kaons is suppressed by 3 orders of magnitude in comparison to the branching ratio of the production from $B$ mesons, but for the considered energies the number of kaons is at least $10^{3}$ times larger than the number of $B$ mesons.

\begin{table}[t!]
\centering
\begin{tabular}{|l|c|c|c|}
 \hline  
 Process  & $\text{BR}(m_S=0)/\theta^2$& Closing mass [GeV] & Appendix\\
 \hline
 $K^\pm\to S\pi^{\pm}$ & $1.7 \cdot 10^{-3}$ & $0.350$ & \ref{app:pseudoscalar}\\ \hline
 $K^0_L\to S\pi^0$ & $7 \cdot 10^{-3}$ & $0.360$ & \ref{app:pseudoscalar}\\ \hline
  $B^{\pm} \to S K^{\pm}_{1}(1270)$ & $(9.1^{+3.6}_{-4.0})\cdot 10^{-1}$ & 3.82 & \ref{app:pseudovector}\\ \hline
  $B^{\pm} \to S K^{*,\pm}_{0}(700)$ & $7.6\cdot 10^{-1}$ & 4.27 & \ref{app:scalar}\\ \hline
  $B^{\pm} \to S K^{*,\pm}(892)$ & $(4.7^{+0.9}_{-0.8})\cdot 10^{-1}$ & 4.29 & \ref{app:vector}\\ \hline
  $B^{\pm} \to S K^{\pm}$ & $(4.3^{+1.1}_{-1.0})\cdot 10^{-1}$ & 4.79 & \ref{app:pseudoscalar} \\ \hline
  $B^{\pm} \to S K^{*,\pm}_{2}(1430)$ & $3.0\cdot 10^{-1}$ & 3.85 & \ref{app:tensor}\\ \hline
  $B^{\pm} \to S K^{*,\pm}(1410)$ & $(2.1^{+0.6}_{-1.1})\cdot 10^{-1}$ & 3.57 & \ref{app:vector}\\ \hline
  $B^{\pm} \to S K^{*,\pm}(1680)$ & $(1.3^{+0.5}_{-0.4})\cdot 10^{-1}$ & 3.26 & \ref{app:vector}\\ \hline
  $B^{\pm} \to S K^{*,\pm}_{0}(1430)$ & $8.1\cdot 10^{-2}$ & 3.82 & \ref{app:scalar}\\ \hline
  $B^{\pm} \to S K^{\pm}_{1}(1400)$ & $(1.6^{+0.6}_{-1.1})\cdot 10^{-2}$ & 2.28 &\ref{app:pseudovector}\\ \hline
  $B^{\pm} \to S \pi^{\pm}$ & $(1.3^{+0.3}_{-0.3})\cdot 10^{-2}$ & 5.14 & \ref{app:pseudoscalar}\\ \hline
\end{tabular}
    \caption{Properties of the main production channels of a scalar $S$ from kaons and $B$ mesons through the mixing with the Higgs boson. \textit{First column}: decay channels; \textit{Second column}: branching ratios of 2-body meson decays evaluated at $m_{S}=0$ using formula~\eqref{eq:productionBRmesons} and normalized by $\theta^2$. For $B$ mesons the numerical values are given for $B^{\pm}$ mesons; in the case of $B^0$ meson all the given branching ratios should be multiplied by a factor of $0.93$ that comes from the difference in total decay widths of $B^{\pm}$ and $B^0$ mesons~\cite{Tanabashi:2018oca}. Uncertainties (where available) follow from uncertainties in meson transition form-factors; \textit{Third column}: effective closing mass, i.e. a mass of a scalar at which the branching ratio of the channel decreases by a factor of $10$; \textit{Fourth column}: a reference to the appendix with details about form-factors used.}
    \label{tab:BR}
\end{table}

\begin{figure}[h!]
    \centering
    \includegraphics[width=0.8\textwidth]{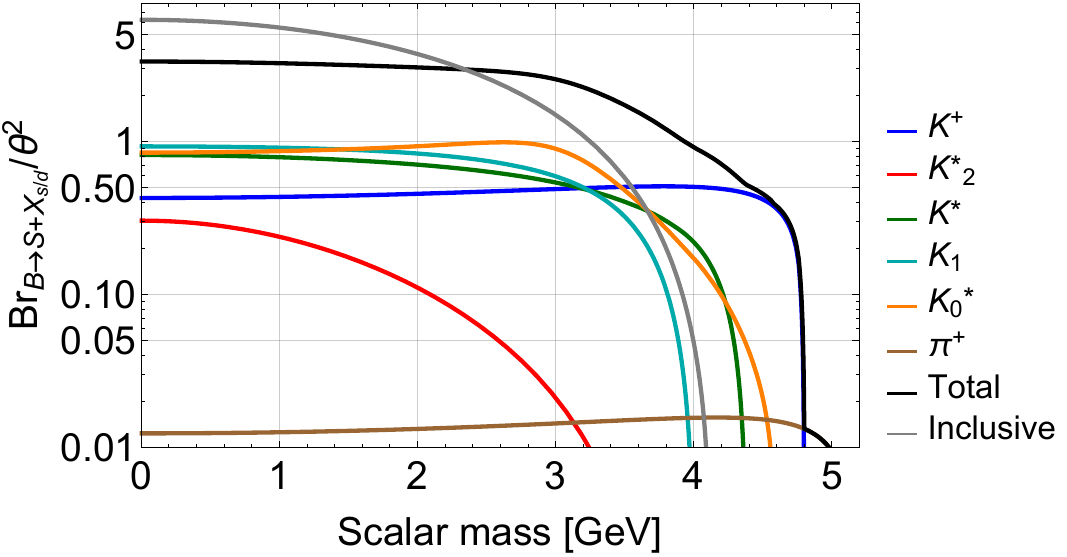}
    \caption{Branching ratios of the 2-body decays $B^{+}\to S X_{s/d}$, where $X_{q}$ is a hadron that contains a quark $q$. By the $K^{*}$ channel, we denote the sum of the branching ratios for $K^{*}(892)$, $K^{*}(1410)$, $K^{*}(1680)$ final states, by $K^{*}_{0}$ -- for $K^{*}_{0}(700)$, $K_{0}^{*}(1430)$, and by $K_{1}$ -- for $K_{1}(1270)$, $K_{1}(1400)$. The ``Inclusive'' line corresponds to the branching ratio~\eqref{eq:prodBRinclusiveB} obtained using the free quark model.}
    \label{fig:plot-b-production}
\end{figure}

\begin{figure}[h!]
    \centering
    \includegraphics[width=0.49\textwidth]{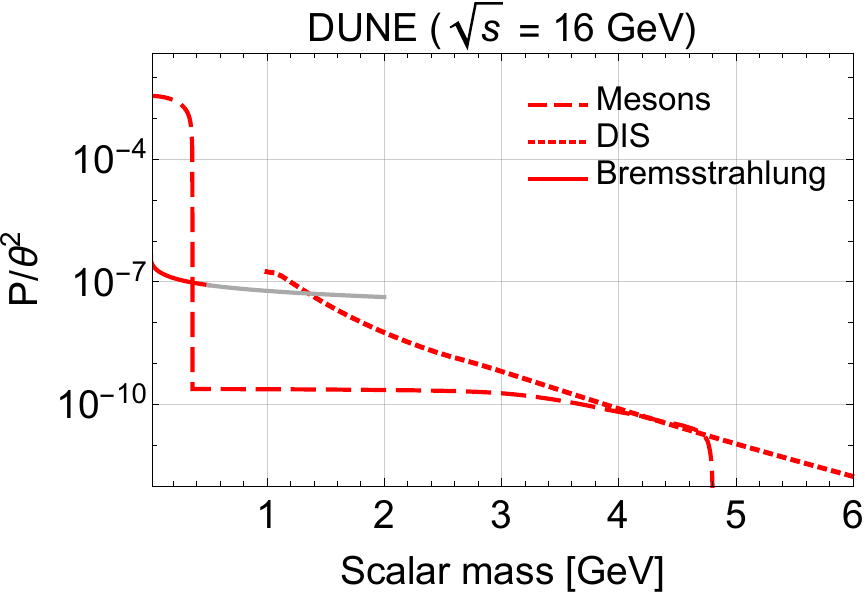}~\includegraphics[width=0.49\textwidth]{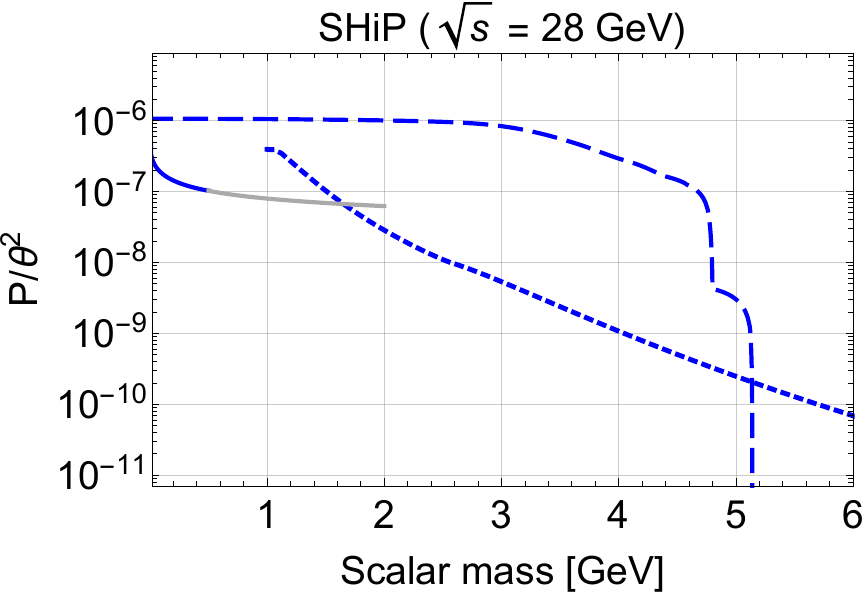} \\
    \includegraphics[width=0.49\textwidth]{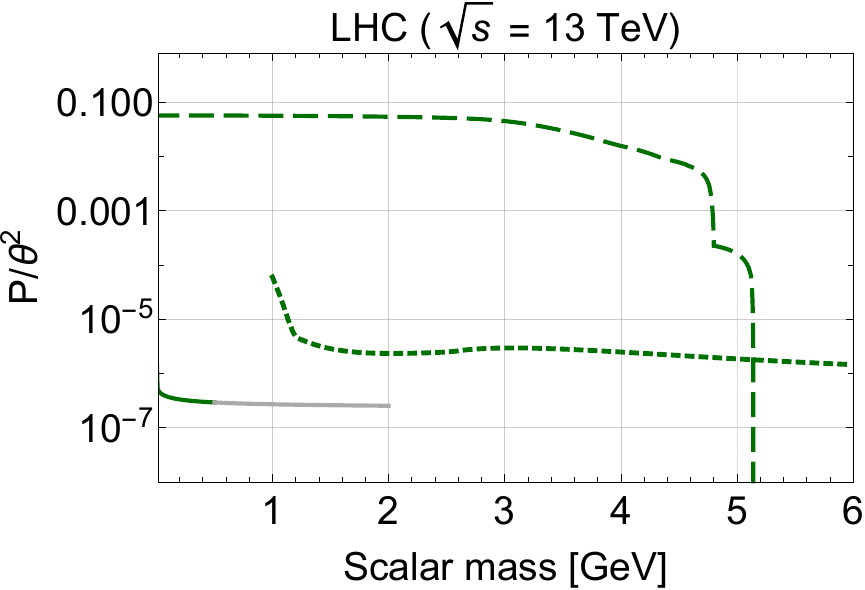}
    \caption{The probabilities of a scalar production in proton bremsstrahlung process~\eqref{eq:Pbrem} (solid lines), DIS process~\eqref{eq:DISsigma} (dotted lines) and decays of $B$ mesons (dashed lines) versus the scalar mass. ``LHC'', ``SHiP'' and ``DUNE'' denote correspondingly results for the proton-proton center-of-mass energies $\sqrt{s_{pp}} = 13\text{ TeV}$, $\sqrt{s_{pp}} = 28\text{ GeV}$ and $\sqrt{s_{pp}} = 16\text{ GeV}$. The gray line corresponds to the extrapolation of the bremsstrahlung production probability assuming unit value of the proton elastic form-factor, see text for details.}
    \label{fig:dis-vs-b-meson}
\end{figure}
For scalar masses $m_K-m_\pi < m_S < m_B$ the main scalar production channel from hadrons is the production from $B$ mesons. Inclusive estimate at the quark level, that we made above (see Eq.~\eqref{eq:PB}), contains an unknown QCD uncertainty and completely breaks down for scalar masses $m_{b} - m_{S} \simeq \Lambda_{\text{QCD}}$.
Below we discuss therefore decays of different mesons containing the b quark
$B \to X_{s/d}S$. We consider kaon and its resonances as the final states $X_{s}$:
\begin{itemize}
    \item Pseudoscalar meson $K$;
    \item Scalar mesons $K_{0}^{*}(700), K_{0}^{*}(1430)$ (here assuming that  $K_{0}^{*}(700)$ is a di-quark state);
    \item Vector mesons $K^{*}(892), K^{*}(1410), K^{*}(1680)$;
    \item Axial-vector mesons $K_{1}(1270), K_{1}(1400)$;
    \item Tensor meson $K_{2}^{*}(1430)$.
\end{itemize}
We also consider the meson $X_{d} = \pi$. Although the rate of the corresponding process $B\to \pi S$ is suppressed in comparison to the rate of $B \to X_{s}S$, it may be important since it has the largest kinematic threshold $m_{S}\lesssim m_{B}-m_{\pi}$.

We calculate the branching ratios $\text{BR}(B^{+}\to X_{s/d}S)$ at $m_{S} \ll m_{K}$ using Eq.~\eqref{eq:productionBRmesons} and give the results in Table~\ref{tab:BR}. The main uncertainty of this approach is related to form factors describing meson transitions $X_{Q_{i}} \to X'_{Q_{j}}$, see Appendix~\ref{sec:hadronic-form-factors} for details. They are calculated theoretically using approaches of light cone fum rules and covariant quark model, and indirectly fixed using experimental data on rare mesons decays~\cite{Ball:2004rg,Ball:2004ye,Cheng:2010yd,Sun:2010nv,Lu:2011jm}. The errors given in Table~\ref{tab:BR} result from uncertainties in the meson transition form-factors $F_{BX_{s/d}}$ (see Appendix~\ref{sec:hadronic-form-factors}). Since $F_{BX_{s/d}}$ are the same for $B^{+}$ and $B^{0}$ mesons, the branching ratios $\text{BR}(B^{0}\to X'^{0}S)$ differ from $\text{BR}(B^{0}\to X_{s/d}^{0}S)$ only by the factor $\Gamma_{B^{+}}/\Gamma_{B^{0}} \approx 0.93$.

The values of the branching ratios for the processes $B \to KS$, $B \to K^{*}S$ are found to be in good agreement with results from the literature~\cite{Batell:2009jf,Winkler:2018qyg}. We conclude that the most efficient production channels of light scalars with $m_{S}\lesssim 3\text{ GeV}$ are decays $B \to K_{0}^{*}S, \ B\to K_{1}S$ and $B\to K^{*}S$; the channel $B \to KS$, considered previously in the literature, is sub-dominant.

Summing over all final $K$ states, in the limit $m_{S}\ll m_{B}$ for the total branching ratio we have
\begin{equation}
\label{eq:total-exclusive}
    \text{BR}(B \to SX_{s}) \equiv \sum_{X_{s}}\text{BR}(B\to SX_{s}) \approx 3.3^{+0.8}_{-0.7}\ \theta^{2}.
\end{equation} 
Using the estimate~\eqref{eq:PB}, for the ratio of the central value of the branching ratio~\eqref{eq:total-exclusive} and the inclusive branching ratio at $m_{S} \ll m_{B}$ we find
\begin{equation}
     \text{BR}(B \to SX_{s})/ \text{BR}_{\text{incl}}(B \to SX_{s}) \approx 0.5.
     \label{eq:exclusive-to-inclusive}
\end{equation}
Provided that the inclusive estimate of the branching ration has a large uncertainty, we believe that Eq.~\eqref{eq:exclusive-to-inclusive} suggests that we have taken into account all main decay channels of this type.
 
Our results for decays of $B$ mesons are summarized in Table~\ref{tab:BR} and Fig.~\ref{fig:plot-b-production}. We have found that the channels with $K^{*}$, $K_{0}^{*}$ and $K_{1}$ give the main contribution to the production branching ratio for small scalar masses $m_{S} \lesssim 3\text{ GeV}$, while for larger masses the main channel is the decay $B^{\pm/0}\to K^{\pm/0}+S$. 

The comparison between the probability of the production from mesons and our estimates for bremsstrahlung and DIS for three center-of-mass energies are shown in Fig.~\ref{fig:dis-vs-b-meson}. In order to indicate an uncertainty because of the unknown proton-scalar elastic form-factor, we show the production probability in bremsstrahlung process for the scalar masses $m_{S}>0.5\text{ GeV}$ in gray (see, however, Fig.~\ref{fig:pp-cross-section}, where we show the bremsstrahlung probability with the form-factor from~\cite{Batell:2020vqn,Foroughi-Abari:2021zbm}).

\subsection{Quartic coupling}
\label{sec:quartic_coupling}

\begin{figure}[t!]
    \centering
    \subfloat[]{\includegraphics[width=0.4\textwidth]{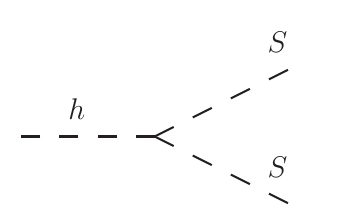}}
    \\
    \subfloat[]{\raisebox{-0.55\height}{\includegraphics[width=0.52\textwidth]{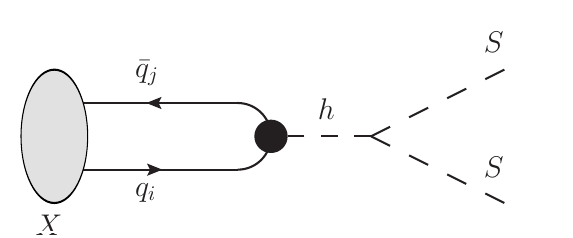}}~\raisebox{-0.5\height}{\includegraphics[width=0.46\textwidth]{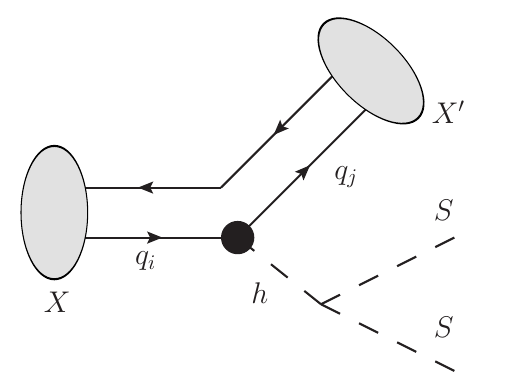}}}
    \caption{Diagrams for the production channels of a scalar $S$ through a quartic coupling: decay of the Higgs boson (a), decays of mesons (b).}
    \label{fig:Sproduction-quartic}
\end{figure}

Above we discussed the production of scalars only through the mixing with the Higgs boson. One more interaction term in the Lagrangian~\eqref{g01},
\begin{equation}
    \mathcal{L}_{\text{quartic}} = \frac{\alpha}{2} S^2 h,
    \label{eq:quartic-coupling-1}
\end{equation}
(the so-called  "quartic coupling" that  originates from the term $S^2 H^{\dagger}H$ in the Lagrangian~\eqref{eq:L1}) affects the production of scalars from decays of mesons and  opens an additional production channel -  production from Higgs boson decays, see Fig.~\ref{fig:Sproduction-quartic}.

\textbf{The production from the Higgs boson (case (a))} can be important for high-energy experiments like LHC. The branching ratio is
\begin{equation}
    \text{BR}(h\to S S) = \frac{\alpha^2|\bm{p}_S|}{16\pi m_h^2 \Gamma_h} \approx 2.0\cdot 10^{-2}\left(\frac{\alpha}{1\text{ GeV}}\right)^{2}\sqrt{1 - \frac{4m_{S}^{2}}{m_{h}^{2}}},
    \label{eq:BR-h-to-SS}
\end{equation}
where $\bm{p}_S$ is a momentum of a scalar in the rest frame of the Higgs boson and we used the SM decay width of the Higgs boson $\Gamma_h \approx 4$~MeV~\cite{Denner:2011mq}. If the decay length of the scalar is large enough $c\gamma\tau_S \gtrsim 1$~m this decay channel manifests itself at ATLAS and CMS experiments as an invisible Higgs boson decay. The invisible Higgs decay is constrained at CMS~\cite{Sirunyan:2018owy}, the $2\sigma$ upper bound is
\begin{equation}
    \text{BR}(h \to \text{invis.}) < 0.19.
    \label{eq:higgs_inv}
\end{equation}
This puts an upper bound $\alpha< 3.1$~GeV for the scalar masses $m_S < m_h/2$.

The  production probability $P_{h\to SS} = \chi_h \times \text{BR}(h\to S S)$, where $\chi_h$ is a production fraction of the Higgs bosons. Comparing with the production from $B$ mesons for a scalar mass below the $B$ threshold estimated by the  inclusive production~\eqref{eq:PB} we get
\begin{equation}
    \frac{P_{h\to SS}}{P_B} \sim 10^{-3} \frac{1}{\theta^2} \left(\frac{\alpha}{1\text{ GeV}}\right)^{2} \frac{\chi_h}{\chi_{b\bar{b}}}
    \sim 10^{-10} \frac{1}{\theta^2} \left(\frac{\alpha}{1\text{ GeV}}\right)^{2},
    \label{eq:Pss-to-PB}
\end{equation}
where we used $\chi_{h}\sim 10^{-9}$ for the LHC energy~\cite{Heinemeyer:2013tqa} and $\chi_{b\bar{b}} \sim 10^{-2}$ (see Table~\ref{tab:meson-amounts}).

Also, \textbf{the quartic coupling generates new channels of scalar production in meson decays (case (b))}.
In addition to the process $X\to X'S$ shown in Fig.~\ref{fig:Sproduction} (c) the quartic coupling enables also additional processes $X\to SS$ and  $X\to X'SS$ shown in Fig.~\ref{fig:Sproduction-quartic} (b) ~\cite{Bird:2004ts,Bird:2006jd,Kim:2009qc,He:2010nt,Batell:2009jf,Badin:2010uh}.

First, let us make a simple comparison between the branching ratios for the scalar production from mesons in the case of mixing with the Higgs boson and quartic coupling. Comparing Feynman diagrams in Figs.~\ref{fig:Sproduction} (c) and \ref{fig:Sproduction-quartic} (b) we see that for the case $m_S\ll m_X$
\begin{equation}
    \frac{\text{BR}(X\to X'SS)}{\text{BR}(X\to X'S)} \sim \frac{\alpha^2 m_X^2}{\theta^2 m_h^4} \sim 10^{-9} \frac{1}{\theta^2} \left(\frac{\alpha}{1\text{ GeV}}\right)^{2}
    \left(\frac{m_X}{1\text{ GeV}}\right)^{2},
    \label{eq:quartic-to-mixing1}
\end{equation}
\begin{equation}
    \frac{\text{BR}(X\to SS)}{\text{BR}(X\to X'S)} \sim \frac{\alpha^2 f_X^2}{\theta^2 m_h^4} \sim 10^{-9} \frac{1}{\theta^2} \left(\frac{\alpha}{1\text{ GeV}}\right)^{2}
    \left(\frac{f_X}{1\text{ GeV}}\right)^{2},
    \label{eq:quartic-to-mixing2}
\end{equation}
where $f_X$ is a meson's decay constant (see Appendix~\ref{sec:quartic-coupling}). 

The channel $X\to X'SS$ is very similar to the channel $X\to X'S$ from Fig.~\ref{fig:Sproduction} (c). By the same reasons, this process is strongly suppressed for $D$-mesons and is efficient only for kaons and  $B$ mesons.
 
The decay $X\to SS$ is possible only for $K^0$, $D^0$, $B^0$ and $B^0_s$ due to conservation of charges. The production from $D^0$ mesons is strongly suppressed by the same reason as above (small Yukawa constant and CKM elements in the effective interaction, see Appendix~\ref{sec:fcnc}).

\begin{figure}[t!]
    \centering
    \includegraphics[width=\textwidth]{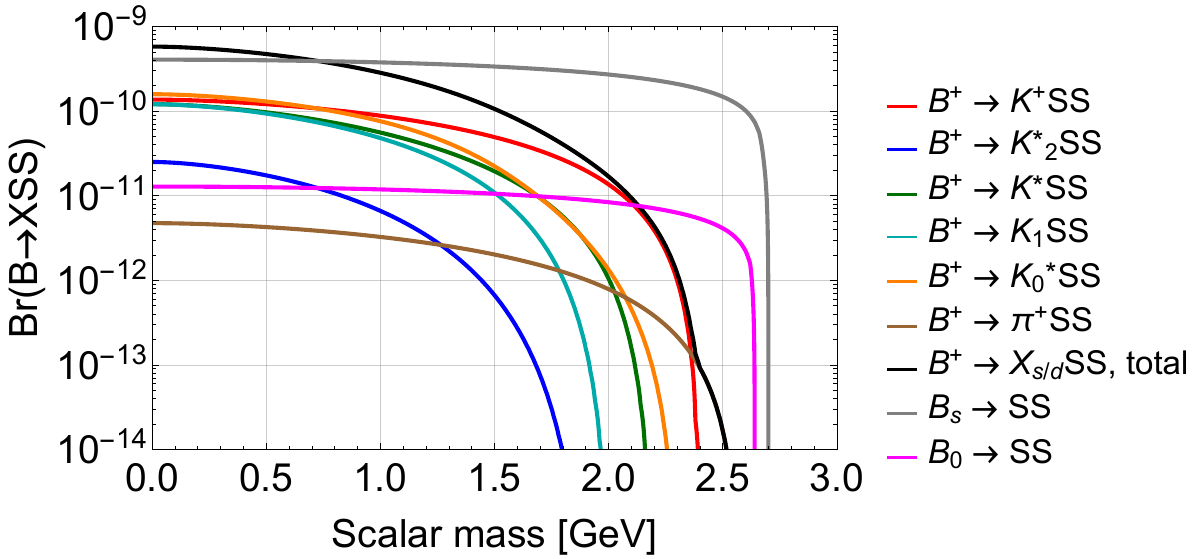}
    \caption{Branching ratios of decays $B^{+}\to SS X_{s/d}$ and $B \to SS$, where $X_{q}$ is a hadron that contains a quark $q$, versus the scalar mass. By the $K^{*}$ channel, we denote the sum of the branching ratios for $K^{*}(892)$, $K^{*}(1410)$, $K^{*}(1680)$ final states, by $K^{*}_{0}$ -- for $K^{*}_{0}(700)$, $K_{0}^{*}(1430)$, and by $K_{1}$ -- for $K_{1}(1270)$, $K_{1}(1400)$. The values of the branching ratios are given at $\alpha = 1\text{ GeV}$.}
    \label{fig:plot-b-production-quartic}
\end{figure}

Our results for the branching ratio of the scalar production from mesons decays in the case of the quartic coupling are presented in Table~\ref{tab:b-production-quartic} and in  Fig.~\ref{fig:plot-b-production-quartic}, the formulas for the branching ratios and details of calculations are given in Appendix~\ref{sec:quartic-coupling}. The results are shown for the value of coupling constant $\alpha = 1$~GeV which corresponds to the Higgs boson branching ratio $\text{BR}(h\to SS) \approx 0.02$ (see Eq.~\eqref{eq:BR-h-to-SS}).
\begin{table}[t!]
\centering
\begin{tabular}{|l|c|c|c|}
 \hline  
 Process  & $\text{BR}(m_S=0)$& Closing mass [GeV] & Appendix \\
 \hline
 $K^0_L\to SS$ & $4.4\cdot 10^{-13}$ & $0.252$ & \ref{sec:quartic-coupling} \\ \hline
 $K^0_L\to SS\pi^0$ & $6.6\cdot 10^{-15}$ & $0.140$ & \ref{app:pseudoscalar} \\ \hline
 $K^\pm\to SS\pi^{\pm}$ & $1.4\cdot 10^{-15}$ & $0.136$ & \ref{app:pseudoscalar}\\ \hline
 $K^{0}_{S}\to SS$ & $7.8\cdot 10^{-16}$ & $0.252$ & \ref{sec:quartic-coupling} \\ \hline
 $K^0_S\to SS\pi^0$ & $1.2\cdot 10^{-17}$ & $0.140$ & \ref{app:pseudoscalar}\\ \hline
  $B_{s}\to SS$ & $4.0\cdot 10^{-10}$ & $2.670$ & \ref{sec:quartic-coupling} \\ \hline
 $B^{\pm} \to SS K^{\pm}$ & $1.4\cdot 10^{-10}$ & $1.998$ &\ref{app:pseudoscalar} \\ \hline
 $B^{\pm} \to SS K^{*,\pm}_{0}(700)$ & $1.2\cdot 10^{-10}$ & $1.621$ &\ref{app:scalar}\\ \hline
    $B^{\pm} \to SS K^{\pm}_{1}(1270)$ & $(1.2^{+0.5}_{-0.5})\cdot 10^{-10}$ & $1.478$ & \ref{app:pseudovector}\\ \hline
  $B^{\pm} \to SS K^{*,\pm}(892)$ & $9.1\cdot 10^{-11}$ & $1.701$ & \ref{app:vector}\\ \hline
  $B^{\pm} \to SS K^{*,\pm}_{0}(1430)$ & $3.5\cdot 10^{-11}$ & $1.621$ & \ref{app:scalar}\\ \hline
  $B^{\pm} \to SS K^{*,\pm}(1410)$ & $(1.9^{+0.6}_{-0.5})\cdot 10^{-11}$ & $1.358$ & \ref{app:vector}\\ \hline
  $B^{\pm} \to SS K^{*,\pm}_{2}(1430)$ & $2.5\cdot 10^{-11}$ & $1.499$ & \ref{app:tensor}\\ \hline
  $B_{0}\to SS$ & $1.1\cdot 10^{-11}$ & $2.624$ & \ref{sec:quartic-coupling}\\ \hline
 $B^{\pm} \to SS K^{*,\pm}(1680)$ & $(9.9^{+0.4}_{-0.3})\cdot 10^{-12}$ & $1.240$ & \ref{app:vector}\\ \hline
  $B^{\pm} \to SS \pi^{\pm}$ & $(4.7^{+1.2}_{-1.1})\cdot 10^{-12}$ & $2.149$ & \ref{app:pseudoscalar}\\ \hline
  $B^{\pm} \to SS K^{\pm}_{1}(1400)$ & $(7.3^{+0.3}_{-0.3})\cdot 10^{-13}$ & $1.545$ &\ref{app:pseudovector}\\ \hline
\end{tabular}
    \caption{Properties of the main production channels of a scalar $S$ from kaons and $B$ mesons through quartic coupling~\eqref{eq:quartic-coupling-1}. \textit{First column}: decay channels; \textit{Second column}: branching ratios of 2-body meson decays evaluated at $m_{S}=0$ and $\alpha = 1\text{ GeV}$, see Eqs.~\eqref{eq:bstossbranching},~\eqref{eq:br-quartic-3-body}. For $B$ mesons the numerical values are given for $B^{\pm}$ mesons; in the case of decays of $B^0$ mesons, all the given branching ratios should be multiplied by a factor $0.93$ that comes from the difference in total decay widths of $B^{\pm}$ and $B^0$ mesons~\cite{Tanabashi:2018oca}. Uncertainties (where available) follow from uncertainties in meson transition form-factors; \textit{Third column}: effective closing mass, i.e. a mass of a scalar at which the branching ratio of the channel decreases by a factor of $10$. \textit{Fourth column}: a reference to appendix with details about form-factors used.}
    \label{tab:b-production-quartic}
\end{table}

\section{Scalar decays}

The main decay channels of the scalar are decays into photons, leptons and hadrons, see Appendix~\ref{sec:decay}. In the mass range $m_{S} \lesssim 2m_{\pi}$ the scalar decays into photons, electrons and muons, see Appendix~\ref{sec:decayleptonphoton}. 

\begin{figure}[h!]
  \centering
    \includegraphics[width=0.55\textwidth]{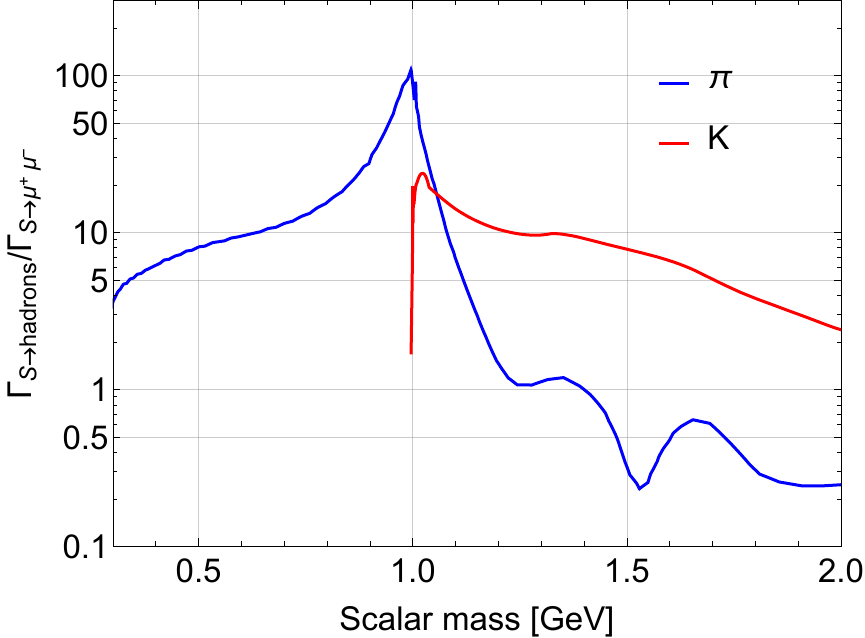}
    \caption{The ratio of the decay widths of a light scalar into pions, kaons and into muons obtained in~\cite{Winkler:2018qyg}. We summed over all final meson states, i.e. for decays into pions we summed over $\pi^{+}\pi^{-}, \pi^{0}\pi^{0}$. A peak in the decay width corresponding to the channel $S\to \pi \pi$ around $m_{S} \simeq 1\text{ GeV}$ is caused by the narrow $f_{0}(980)$ resonance.}
    \label{fig:Winkler-Br}
\end{figure}

\begin{figure}[h!]
  \centering
  \begin{minipage}[b]{0.48\textwidth}
    \includegraphics[width=\textwidth]{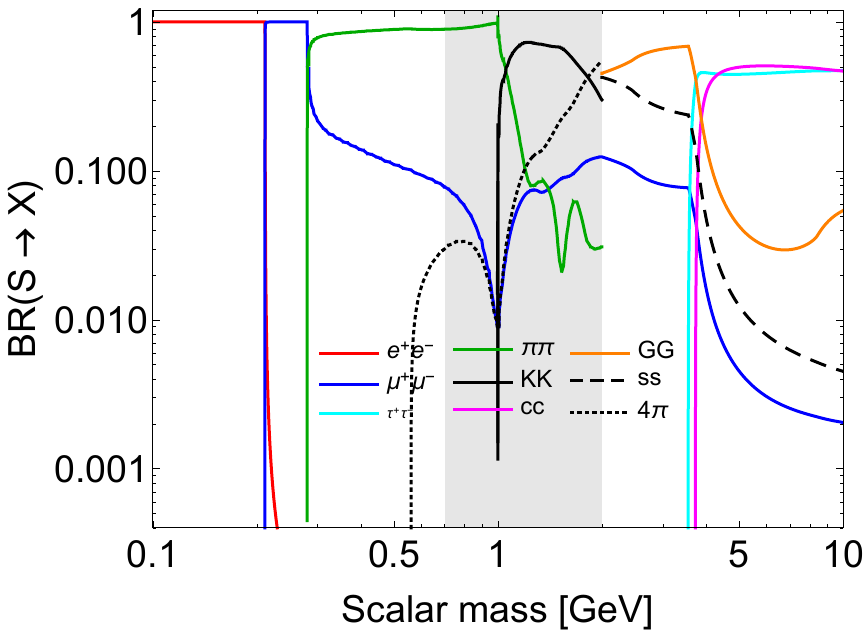}
  \end{minipage}
  \hfill
  \begin{minipage}[b]{0.48\textwidth}
    \includegraphics[width=\textwidth]{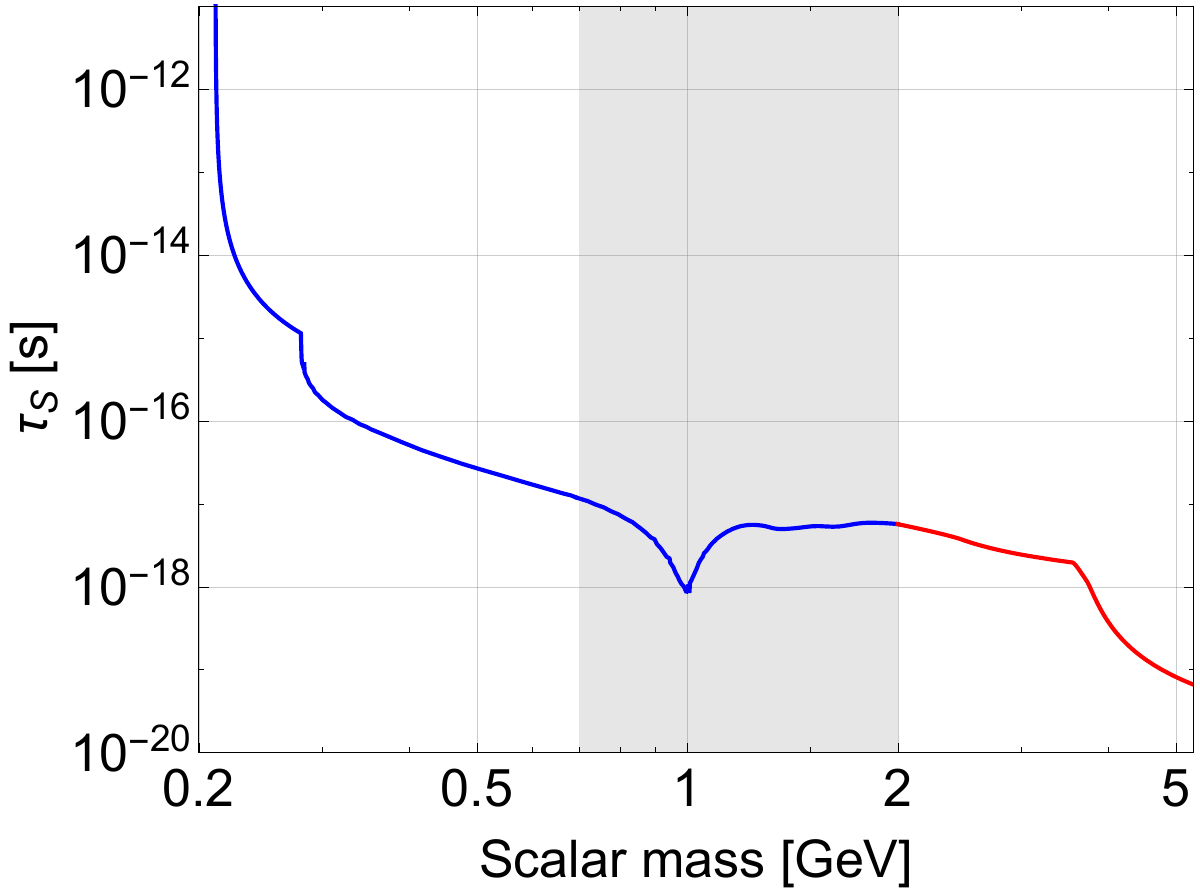}
  \end{minipage}
  \caption{Left panel: branching ratios of decays of a scalar $S$ as a function of its mass. For decays into hadrons up to $m_{S} =2\text{ GeV}$ we used results from~\cite{Winkler:2018qyg}, while for decays into hadrons in the mass range $m_{S} > 2\text{ GeV}$ we used perturbative decays into quarks and gluons (see Sec.~\ref{sec:decay-perturbative}). In order to match these two regimes, we added a toy-model contribution to the total decay width that imitates multi-meson decay channels, see Eq.~\eqref{eq:Gmultimeson}. Right panel: the lifetime of a scalar $S$ as a function of its mass with the mixing angle $\theta^2 = 1$. Solid blue line denotes the lifetime calculated using decays into leptons, kaons and pions from~\cite{Winkler:2018qyg} and fictitious multi-meson channel, see Eq.~\eqref{eq:Gmultimeson}, while solid red line -- the lifetime obtained using decays into quarks and gluons within the framework of perturbative QCD. The filled gray regions on the plot correspond to the domain of the scalar masses for which there are significant theoretical uncertainties in hadronic decays.}
  \label{fig:decays}
\end{figure}

For masses small enough compared to the cutoff $\Lambda_{\text{QCD}}\approx 1~$GeV, ChPT (Chiral Perturbation Theory) can be used to predict the decay width into pions~\cite{Voloshin:1985tc}. For masses of order $m_{S} \gtrsim \Lambda_{\text{QCD}}$ a method making use of dispersion relations was employed in~\cite{Raby:1988qf,Truong:1989my,Donoghue:1990xh} to compute hadronic decay rates. As it was pointed out in~\cite{Monin:2018lee} and later in~\cite{Bezrukov:2018yvd}, the reliability of the dispersion relation method is questionable for $m_{S}\gtrsim 1\text{ GeV}$ because of lack of the experimental data on meson scattering at high energies and unknown contribution of scalar hadronic resonances, which provides significant theoretical uncertainties. To have a concrete benchmark -- although in the light of the above the result should be taken with a grain of salt -- we use the decay width into pions and kaons from~\cite{Winkler:2018qyg}, see Fig.~\ref{fig:Winkler-Br}, which combines the next-to-leading order ChPT with the analysis of dispersion relations for the recent experimental data. For the ratio of the decay widths into neutral and charged mesons we have
\begin{equation}
    \Gamma_{S\to \pi^{0}\pi^{0}}/\Gamma_{S\to \pi^{+}\pi^{-}} = 1/2, \quad \Gamma_{S\to K^{0}\bar{K}^{0}}/\Gamma_{S\to K^{+}K^{-}} = 1. 
\end{equation}
For scalar masses above $f_0(1370)$ the channel $S\to 4\pi$ becomes important and should be taken into account~\cite{Moussallam:1999aq}. The decay width of this channel is currently not known; its contribution can be approximated by a toy-model formula~\cite{Winkler:2018qyg}
\begin{equation}
\Gamma_{\text{multi-meson}} = C\theta^{2}m_{S}^{3}\beta_{2\pi}, \quad \beta_{2\pi} = \sqrt{1-(4m_{\pi})^{2}/m_{S}^{2}},
\label{eq:Gmultimeson}
\end{equation}
where a dimensional constant $C$ is chosen so that the total decay width is continuous at large $m_S$ that will be described by perturbation QCD, see Fig.~\ref{fig:decays}.

For $m_S \geq \Lambda_{S}^{\text{pert}} \simeq 2-4$~GeV hadronic decays of a scalar can be described perturbatively using decays into quarks and gluons, see Appendix~\ref{sec:decay-perturbative}. Currently, the value of $\Lambda_{S}^{\text{pert}}$ is not known because of lack of knowledge about scalar resonances which can mix with $S$ and enhance the scalar decay width. In~\cite{Winkler:2018qyg} the value of $\Lambda_{S}^{\text{pert}}$ is set to $2\text{ GeV}$, in~\cite{Bezrukov:2009yw} it is $\Lambda_{S}^{\text{pert}} = 2.5\text{ GeV}$, while in~\cite{Evans:2017lvd} its value is $\Lambda_{S}^{\text{pert}} = 2m_{D}$. This scale certainly should above the mass of the heaviest known scalar resonance $f_{0}(1710)$, so in this work we choose $\Lambda_{S}^{\text{pert}} = 2$~GeV.

The summary of branching ratios of various decay channels of the scalar and the total lifetime of the scalar is shown in Fig.~\ref{fig:decays}.

\begin{small}
\LTcapwidth=\textwidth
\begin{longtable}{|c|c|c|c|c|c|}
  \caption[Table of the relevant scalar decay channels]{\small
  Relevant scalar decay channels. Only channels with the branching ratio above $1\%$ covering the scalar mass range up to $10$~GeV are shown. The gray line corresponds to the fake multi-meson channel, see discussion in text.\newline
  \emph{Columns}: (1) the scalar decay channel.
  (2) The scalar mass at which the channel opens;
  (3) The scalar mass starting from which the channel becomes relevant (contributes larger than $10\%$);
  (4) The scalar mass above which the channel contributes less than $10\%$; ``---'' means that the channel is still relevant at $m_S=10$~GeV;
  (5) The maximal branching ratio of the channel for $m_S<10$~GeV;
  (6) Reference to the formula (or figure in case of decays into pions and kaons) for the decay width of the channel.
\label{tab:decaychannels}} \\
\hline
  Channel & Opens at& Relevant from & Relevant to & Max BR & Reference \\
  & [MeV]  &  [MeV]  &  [MeV] &  [\%] & in text\\
  \endfirsthead 
\caption[]{(continued)} \\
\hline
Channel & Open, MeV & Rel. from, MeV & Rel. to, MeV & Max BR, \% & Formula \endhead \hline
$S\to 2\gamma$ 
    & 0 & 0 & 1.02 & 100 & (\ref{eq:Sto2gamma}) \\ \hline
$S\to e^+ e^-$ 
    & 1.02 & 1.02 & 212 & $\approx 100$ & (\ref{eq:Gll}) \\ \hline
$S\to \mu^+ \mu^-$ 
    & 211 & 211 and 1668 & 564 and 2527 & $\approx 100$ & (\ref{eq:Gll}) \\ \hline
$S\to \pi^+ \pi^-$ 
    & 279 & 280 & 1163 & 65.5 & Fig.~\ref{fig:Winkler-Br} \\ \hline
$S\to 2 \pi^0$ 
    & 270 & 280 & 1163 & 32.8 & Fig.~\ref{fig:Winkler-Br} \\ \hline
$S\to K^+ K^-$ 
    & 987 & 996 & $\Lambda_{S}^{\text{pert}} = 2000$ & 36.8 & Fig.~\ref{fig:Winkler-Br}  \\ \hline
$S\to K^0 \bar{K}^0$ 
    & 995 & 1004 & $\Lambda_{S}^{\text{pert}} = 2000$ & 36.8 & Fig.~\ref{fig:Winkler-Br} \\ \hline
{\color{gray} $S\to 4\pi$}
    & {\color{gray} 550} & {\color{gray} 1210} & {\color{gray} $\Lambda_{S}^{\text{pert}} = 2000$} & {\color{gray} 52.4} &  \eqref{eq:Gmultimeson} \\ \hline
$S\to GG$ 
    & $275$ & $\Lambda_{S}^{\text{pert}} = 2000$ & 4178 & 68.6 & (\ref{eq:decay-gluons}) \\ \hline
$S\to s \bar{s}$ 
    & 990 & $\Lambda_{S}^{\text{pert}} = 2000$ & 3807 & 42.5 & (\ref{eq:decay-quarks}) \\ \hline
$S\to \tau^+\tau^-$ 
    & 3560 & 3608 & --- & 45.7 & (\ref{eq:Gll}) \\ \hline
$S\to c \bar{c}$ 
    & 3740 & 3797 & --- & 50.6 & (\ref{eq:decay-quarks}) \\ \hline
\end{longtable}
\end{small}

\newpage

\section{Conclusion}

In this paper, we have reviewed and revised the phenomenology of the scalar portal, a simple extension of the Standard Model with a scalar $S$ that is not charged under the SM gauge group, for masses of scalar $m_S \lesssim 10$~GeV. 
We considered three examples of experimental setup that correspond to DUNE (with proton-proton center of mass energy $\sqrt{s_{pp}} \approx 16$~GeV), SHiP ($\sqrt{s_{pp}} \approx 28$~GeV) and LHC based experiments ($\sqrt{s_{pp}} = 13$~TeV). 

Interactions of a scalar $S$ with the Standard Model can be induced by the mixing with the Higgs boson and the interaction $S h^2$ (the ``quartic coupling''), see  Lagrangian~\eqref{eq:L1}. The mixing with the Higgs boson is relevant for a scalar production and decay, while the quartic coupling could be important only for the scalar production. 
 
For the scalar production through the mixing with the Higgs boson, we have explicitly compared decays of secondary mesons, proton bremsstrahlung, photon-photon fusion, and deep inelastic scattering. For the energy of the SHiP experiment, the most relevant production channel is the production in decays of secondary mesons, specifically kaons and $B$ mesons. For smaller energies (corresponding in our examples to the DUNE experiment) the situation is more complicated, and direct production channels from $p-p$ collisions (proton bremsstrahlung, deep inelastic scattering) give the main contribution to the production of scalars, see Fig.~\ref{fig:dis-vs-b-meson}.
 
Our results for various channels of the scalar production from mesons via mixing with the Higgs boson are summarized in Table~\ref{tab:BR} and in Fig.~\ref{fig:plot-b-production}. The results for decays $B\to KS$, $B\to K^{*}(892)S$ agree with the references~\cite{Bezrukov:2009yw,Clarke:2013aya,Winkler:2018qyg,Bird:2004ts,Pospelov:2007mp,Batell:2009jf}, while other decay channels have not been studied in these papers.

For the LHC-based experiments, an important contribution to the production of scalars is given by the production in decays of Higgs bosons that may be possible due to non-zero quartic coupling.  This production channel, when allowed by the energy of an experiment, allows to search for scalars that are heavier than $B$ mesons. It may also significantly increase the experimental sensitivity in the region of the lower bound of the sensitivity curve, where production through the mixing with the Higgs boson is less efficient. 

The quartic coupling also gives rise to meson decay channels $X\to SS$ and $X\to X' SS$, that are important for scalar masses $m_S \lesssim m_B/2$. Our results for these channels are presented in Table~\ref{tab:b-production-quartic} and in Fig.~\ref{fig:plot-b-production-quartic}.
 
The description of scalar decays is significantly affected by two theoretical uncertainties: (i) the decay width of a scalar into mesons like $S\to \pi \pi$ and $S \to KK$ (that may be uncertain more than by an order of magnitude for masses of a scalar around 1~GeV) and (ii) the uncertainty in the scale $\Lambda_S^{\text{pert}}$ at which perturbative QCD description can be used. As a benchmark, for decays into mesons we use results of~\cite{Winkler:2018qyg} and choose  $\Lambda_{s}^{\text{pert}} = 2\text{ GeV}$, but we stress that the correct result is not really known for such masses. The main properties of scalar decays are summarized in Table~\ref{tab:decaychannels} and Fig.~\ref{fig:decays}.

\subsection*{Acknowledgements}

We thank O.~Ruchayskiy, F.~Bezrukov, J.~Bluemlein, A.~Manohar, A.~Monin for fruitful discussions and J.-L.~Tastet for careful reading of the manuscript. This project has received funding from the European Research Council (ERC) under the European Union's Horizon 2020 research and innovation program (GA 694896).

\appendix

\section{Effective interactions}
\label{sec:effective-interactions}
\subsection{Photons and gluons}
\label{sec:effective-interactions-gauge-bosons}
\begin{figure}[h!]
    \centering
    \includegraphics[width=0.32\textwidth]{scalar-photon-photon-vertex.pdf}~\includegraphics[width=0.32\textwidth]{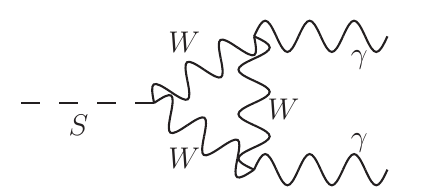}~\includegraphics[width=0.32\textwidth]{scalar-gluon-gluon-vertex.pdf}
    \caption{Diagrams of the interaction of the scalar $S$ with photons and gluons.}
    \label{fig:scalar-gauge-bosons-interactions}
\end{figure}
\begin{figure}[h!]
    \centering
    \includegraphics[width=0.7\textwidth]{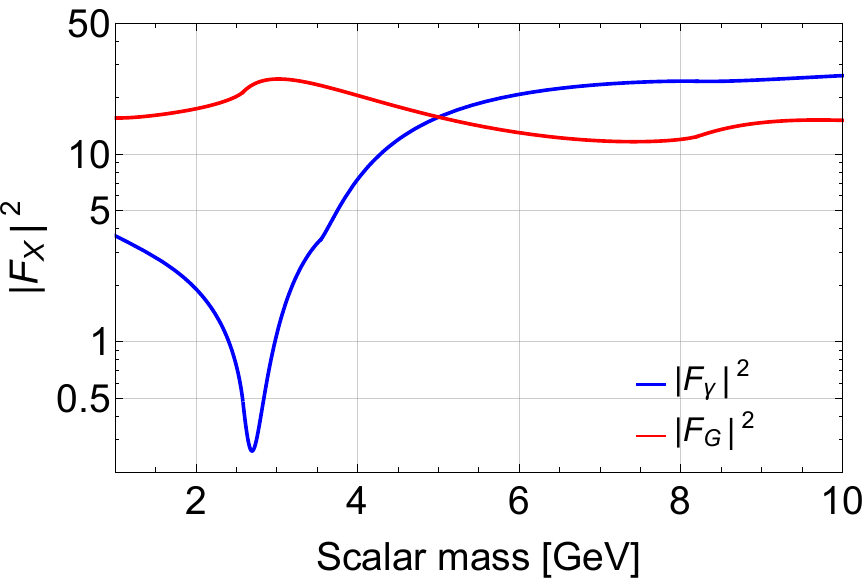}
    \caption{Dependence of the photon and gluon loop factors~\eqref{eq:photon-gluon-loop-factor} on the scalar mass.}
    \label{fig:quark-gluon-loop-factor}
\end{figure}

The effective lagrangian of the interaction of $S$ with photons and gluons is generated by the diagrams~\ref{fig:scalar-gauge-bosons-interactions}. It reads
\begin{equation}
\label{eq:LggS}
    \mathcal{L} = \theta S C_{S\gamma\gamma}\frac{\alpha_{\text{EM}}}{4\pi v}F_{\gamma}F_{\mu\nu}F^{\mu\nu}+\theta SC_{SGG}\frac{\alpha_{s}}{4\pi v}F_{G}\sum_{a}G_{\mu\nu}^{a}G^{\mu\nu,a}.
\end{equation}
Here the effective vertices $F_{\gamma}, F_{G}$ are~\cite{Bezrukov:2009yw,Spira:2016ztx}
\begin{equation}
\label{eq:photon-gluon-loop-factor}
    F_{\gamma} = \sum_{l = e,\mu,\tau}F_{l}+N_{c}\sum_{q}F_{q} +F_{W}, \quad F_{G} = \sum_q F_{q},
\end{equation}
where
\begin{equation}
    F_{f}(l_{f}) = -2l_{f}(1+(1-l_{f})x^{2}(l_{f})), \quad F_{W} = 2+3l_{W}(1+(2-l_{W})x^{2}(l_{W})), \quad l_{X} = 4m_{X}^{2}/m_{S}^{2},
\end{equation}
and
\begin{equation}\label{xnotation}
    x(l) = \begin{cases}\arctan\left(\frac{1}{\sqrt{l-1}}\right), \quad l > 1, \\ \frac{1}{2}\left( \pi + i\ \ln\left[\frac{1+\sqrt{1-l}}{1-\sqrt{1-l}}\right]\right), \quad l<1\end{cases}
\end{equation}
Their behavior in dependence on the scalar mass is shown in Fig.~\ref{fig:quark-gluon-loop-factor}.
The values of the constants $C_{SGG}$ and $C_{S\gamma\gamma}$ vary in the literature. Namely, in~\cite{Bezrukov:2009yw} they are $C_{S\gamma\gamma} = 1, C_{SGG} = 1/\sqrt{8}$. 
From the other side, in~\cite{Spira:2016ztx} predicts $|C_{S\gamma\gamma}| = 1/2, |C_{SGG}| = 1/4$. Calculating the decay branching ratio of the Higgs boson into two photons, we found that the value $C_{S\gamma\gamma} = 1/2$ is consistent with experimental results for the signal strength of the process $p + p \to h +X, \ h \to \gamma\gamma$~\cite{Khachatryan:2016vau}.\footnote{We used the Higgs boson decay width $\Gamma_{h,\text{SM}} = 4\text{ MeV}$.} The gluon loop factor in the triangle diagram~\ref{fig:scalar-gauge-bosons-interactions} differs from the photon loop factor by the factor $\text{tr}[t_{a}t_{a}] = \frac{1}{2}$, where $t_{a}$ is the QCD gauge group generators, and therefore $C_{Sgg} = 1/4$.

\subsection{Nucleons}
\label{sec:effective-interactions-nucleons}
Consider the low-energy interaction Lagrangian between the nucleons $N$ and the scalar: 
\begin{equation}
\label{eq:SNN}
\mathcal{L}_{S NN} = g_{S NN}\theta S\bar{N}N
\end{equation} 
The coupling $g_{SNN}$ is defined as
\begin{equation}
\label{eq:SNN-coupling}
g_{S NN} \equiv \frac{1}{v}\lim_{p\to p'}\langle N(p)|\sum_{q}m_{q}\bar{q}q|N(p')\rangle \equiv \frac{1}{v}\langle N|\sum_{q}m_{q}\bar{q}q|N\rangle,
\end{equation}
where the shorthand notation $\langle N|..|N\rangle \equiv \lim_{p\to p'}\langle N(p)|..|N(p')\rangle$ was used. The applicability of the effective interaction~\eqref{eq:SNN} is $m_{S}^{2}\ll r_{N}^{-2} \simeq 1\text{ GeV}^{2}$. Above this scale, the elastic $SNN$ vertex competes with the inelastic processes on partonic level, and hence it is suppressed.

For energy scales of order of the nucleon mass, the $u,d,s$ quarks are light, while the $c,b,t$ quarks are heavy. Therefore, the latter can contribute to the effective coupling~\eqref{eq:SNN-coupling} only through effective interactions involving the lighs quarks and gluons. The latter can be obtained using the heavy quarks expansion~\cite{Vainshtein:1975sv,Witten:1975bh}. Keeping only the leading $1/m_{q_{\text{heavy}}}$ term, for the effective interaction operator we obtain~\cite{Shifman:1978zn}
\begin{equation}
\label{eq:hqeft}
\sum_{q = c,b,t}m_{q}\bar{q}q \to -n_{\text{heavy}}\cdot \frac{\alpha_{s}}{12\pi}G_{\mu\nu}^{a}G^{\mu\nu,a} + O\left(\frac{1}{m_{q_{\text{heavy}}}^{2}}\right).
\end{equation}
Here $\alpha_{s}$ is the QCD interaction constant evaluated on the scale of the hadronic mass, $G_{\mu\nu}^{a}$ is the gluon strength tensor and $n_{\text{heavy}} = 3$ is the number of the heavy quarks. Therefore, in the leading order of $1/m_{q_{\text{heavy}}}$ expansion the coupling~\eqref{eq:SNN-coupling} takes the form~\cite{Shifman:1978zn,Cheng:1988im}
\begin{equation}
\label{eq:SN1}
g_{SNN} = \frac{\theta}{v}\langle N|\sum_{q = u,d,s}m_{q}\bar{q}q -n_{\text{heavy}}\frac{\alpha_{s}}{12\pi}G_{\mu\nu}^{a}G^{\mu\nu,a}|N\rangle.
\end{equation}

The last expression we can written in terms of the nucleon mass $m_{N}$,
\begin{equation}
m_{N} \equiv \langle N|\theta_{\mu\mu}|N \rangle,     
\end{equation}
where $\theta_{\mu\mu}$ is the trace of the stress-energy tensor in the QCD~\cite{Shifman:1978zn}
\begin{equation}
\label{eq:trace}
\theta_{\mu\mu} = \sum_{q = u,d,s}m_{q}\bar{q}q +\frac{\beta_{s}}{4 \alpha_{s}}G_{\mu\nu}^{a}G^{\mu\nu,a},
\end{equation}
where $\beta_{s}$ is the QCD $\beta$ function,
\begin{equation}
\beta_{s} = -\left(9-\frac{2}{3}n_{\text{heavy}}\right)\frac{\alpha_{s}^{2}}{2\pi},
\end{equation}
in the leading order by $\alpha_s$.
Therefore, we get~\cite{Shifman:1978zn,Cheng:1988im}
\begin{equation}
g_{S NN} =\frac{2}{9}\frac{m_{N}}{v}\left( 1 + \frac{7}{2}\sum_{q = u,d,s}\frac{m_{q}}{m_{N}}\langle N|\bar{q}q|N\rangle\right).
\end{equation}
The numerical value is $g_{SNN} \approx 1.2\cdot 10^{-3}$~\cite{Cheng:2012qr}.

In order to incorporate effects of non-zero momentum transfer $q^{2}$ in the $SNN$ vertex, we need to take into account an scalar nucleon form-factor $F_{N,S}(q^{2})$:
\begin{equation}
    g_{SNN} \to g_{SNN}F_{N,S}(q^{2}), \quad F_{N,S}(0) = 1.
\end{equation}
From general grounds, we expect that the form-factor $F_{N,S}(q^{2})$ incorporates a mixing with scalar $J^{PC}= 0^{++}$ resonances, which causes the resonant enhancement of the production at $q^{2} = m_{f_{0}}^{2}$. A model for the form-factor is discussed in the works~\cite{Batell:2020vqn,Foroughi-Abari:2021zbm}. According to it, the scalar incorporates mixing through a sum of Breit-Wigner components of the lightest scalar resonances $f_{0}(500),f_{0}(980),f_{0}(1370)$:
\begin{equation}
    F_{N,S}(q^{2})=\sum_{f_{0}}\frac{f_{\phi}m_{\phi}^{2}}{m_{\phi}^{2}-q^{2}-im_{f_{0}}\Gamma_{f_{0}}}
    \label{eq:scalar-nucleon-form-factor}
\end{equation}
The constants are extracted from the criteria $F_{N,S}(0) = 1$ and $F_{N,S}(q^{2}) \propto q^{-4}$ at $q^{2}\gg m_{f_{0}}^{2}$, with the latter following from the quark counting rules. Their central numerical values are $f_{0} = \{0.28,1.8,-0.99\}$. 

\subsection{Flavor changing effective Lagrangian}
\label{sec:fcnc}

\begin{figure}[h!]
\centering
\includegraphics[width=\textwidth]{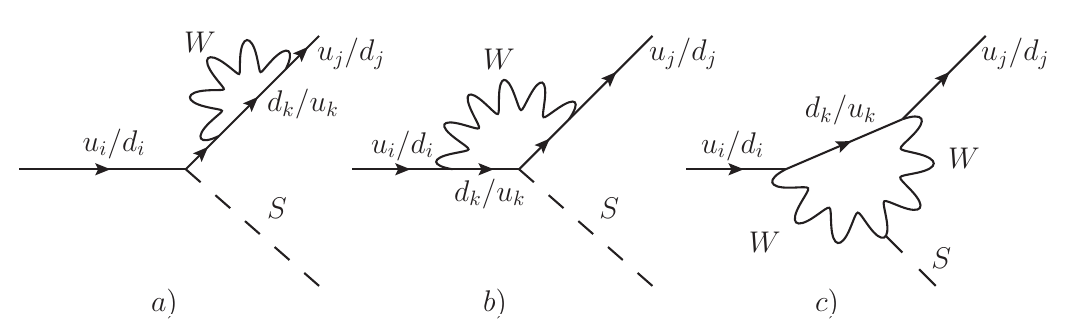} 
\caption{Diagrams of the production of the scalar $S$ in flavor changing quarks transitions in the unitary gauge.}
\label{fig:3diagrams}
\end{figure}

A light scalar $S$ can be produced from a hadron via flavor changing quarks transitions (see diagrams in Fig.~\ref{fig:3diagrams}). The flavor changing amplitude was calculated using different techniques in many papers~\cite{Willey:1982ti,Willey:1986mj,Grzadkowski:1983yp, Leutwyler:1989xj,Haber:1987ua,Chivukula:1988gp}. The corresponding effective Lagrangian of flavor changing quark interactions with the $S$ particle is
\begin{equation}
\label{eq:g05}
\mathcal{L}_{eff}^{Sqq}= \theta\ \frac{S}{v}\sum_{i,j}\xi_{ij}m_{Q_j}\bar{Q}_{i}P_{R}Q_{j}+h.c.,
\end{equation}
where $Q_i$ and $Q_j$ are both upper or lower quarks and $P_R \equiv (1+\gamma_{5})/2$ is a projector on the right chiral state. The effective coupling $\xi_{ij}$ is defined as
\begin{equation}
    \xi_{ij} = \frac{3G_{F}\sqrt{2}}{16\pi^{2}}\sum_{k}V_{ki}^{*}m_{k}^{2}V_{kj},
\end{equation}
where $Q_k$ are the lower quarks if $Q_i$ and $Q_j$ are the upper and vice versa, $V_{ij}$ are the elements of the CKM matrix, and $G_F$ is the Fermi constant. One power of the quark mass in the expression for $\xi$ comes from the $h\bar{q}q$ coupling, while another one comes from the helicity flip on the quark line in the diagrams in Fig.~\ref{fig:3diagrams}. Because of such behavior, the quark transition generated by the Lagrangian~\eqref{eq:g05} is more probable for lower quarks than for upper ones, since the former goes through the virtual top quark. Numerical values of some of $\xi_{ij}$ constants are given in Table~\ref{tab:xi}.
\begin{table}[h]
\centering
\begin{tabular}{|c|c|c|c|c|}
 \hline  
 $\xi_{ij}$  & $\xi_{ds}$ & $\xi_{uc}$ & $\xi_{db}$ & $\xi_{sb}$ \\
 \hline
 Value& $3.3 \cdot 10^{-6}$ & $1.4 \cdot 10^{-9}$  & $7.9 \cdot 10^{-5}$ & $3.6 \cdot 10^{-4}$\\
 \hline
\end{tabular}
    \caption{Numerical values of $\xi_{ij}$ constants in effective Lagrangian~\eqref{eq:g05}.}
    \label{tab:xi}
\end{table}

\section{Scalar production from mesons}
\label{sec:production-hadronic-decays}

In the scalar production from hadron decays, the main contribution comes from the lightest hadrons in each flavor, which are mesons.\footnote{Indeed, if $X$ is the lightest hadron in the family, it could decay \textit{only} through weak interaction, so it has small decay width $\Gamma_X$ (in comparison to hadrons that could decay through electromagnetic or strong interactions). The probability of light scalar production from hadron is inversely proportional to hadron decay width thus the light scalar production from the lightest hadrons is the most efficient.} The list of the main hadron candidates is the following (the information is given in the format ``Hadron name (quark contents, mass in MeV)''):
\begin{itemize}
    \item s-mesons $K^-(s\bar{u}, 494)$, $K^0_{S,L}(s\bar{d}, 498)$;
    \item c-mesons $D^0(c\bar{u},1865)$, $D^+(c\bar{d},1870)$, $D_s(c\bar{s},1968)$, $J/\psi(c\bar{c},3097)$;
    \item b-mesons $B^-(b\bar{u},5279)$, $B^0(b\bar{d},5280)$, 
    $B_s (b\bar{s},5367)$, $B_c (b\bar{c},6276)$, $\Upsilon (b\bar{b},9460)$.
\end{itemize}
The production of a scalar from mesons is possible through the flavor changing neutral current~\ref{sec:fcnc}, so the production from $D_s$, $J/\psi$, $B_s$, $B_c$ and $\Upsilon$ mesons does not have any advantage with respect to the production from $D^0$, $D^+$, $B^-$ and $B^0$, while their amount at any experiment is significantly lower. Therefore, we will discuss below only production from later mesons.

\subsection{Inclusive production}
\label{sec:inclusive-br-calculation}
The decay widths for the processes $Q_{i}\to Q_{j}S$, $Q_{i}\to Q'e\bar{\nu}_{e}$ are
\begin{equation}
    \Gamma_{Q_{i}\to Q_{j} +S} = \frac{|\mathcal{M}_{Q_{i}\to Q_{j}S}|^{2}}{8\pi m_{b}}\frac{|\bm{p}_{S}|}{m_{b}} \approx |\xi_{bs}|^{2}\frac{m_{b}^{3}\left(1-\frac{m_{S}^{2}}{m_{b}^{2}} \right)^{2}}{32 \pi v^{2}}\theta^{2},
\end{equation}
\begin{multline}
    \Gamma_{Q_{i}\to Q_{k} +e+\bar{\nu}_{e}} = \frac{1}{(2\pi)^{3}}\int \limits_{m_{Q_{k}}^{2}}^{m_{Q_{i}}^{2}}ds_{Q_{k}e} \int \limits_{s_{e\nu,\text{min}}}^{s_{e\nu,\text{max}}} ds_{e\nu} \frac{|\mathcal{M}_{Q_{i}\to Q_{k}+e+\bar{\nu}_{e}}|^{2}}{32m_{Q_{i}}^{3}} \approx \\ \approx
    \frac{G_{F}^{2}|V_{Q_{i}Q_{k}}|^{2}m_{Q_{i}}^{5}}{192\pi^{3}}\times f(m_{Q_{k}}/m_{Q_{i}}),
\end{multline}
where $\bm{p}_{S}$ is the $S$ particle momentum at the rest frame of the meson $X$,
\begin{equation}
    |\bm{p}_{S}| = \frac{\sqrt{(m_{X}^{2}-(m_{S}+m_{X'})^{2})((m_{X}^{2}-(m_{S}-m_{X'})^{2})}}{2m_{X}},
\end{equation}
the integration limits are
\begin{equation}
s_{e\nu,\text{min}} = 0, \quad s_{e\nu,\text{max}}= m_{Q_{i}}^{2}+m_{Q_{i}}^{2}-s_{Q_{k}e}-\frac{m_{Q_{i}}^{2}m_{Q_{k}}^{2}}{s_{Q_{k}e}},
\end{equation}
and
\begin{equation}
    f(m_{Q_{k}}/m_{Q_{i}}) = \bigg(1-8\frac{m_{Q_{k}}^{2}}{m_{Q_{i}}^{2}}-24\frac{m_{Q_{k}}^{4}}{m_{Q_{i}}^{4}}\ln\left(\frac{m_{Q_{k}}}{m_{Q_{i}}}\right)+8\frac{m_{Q_{k}}^{6}}{m_{Q_{i}}^{6}}-\frac{m_{Q_{k}}^{8}}{m_{Q_{i}}^{8}} \bigg) \approx 1/2
\end{equation}
is the phase space factor.

\subsection{Scalar production in two-body mesons decays}
\label{sec:meson-two-body-decays}

Let us consider exclusive 2-body decay of a meson
\begin{equation}
    X_{Q_i}\to X'_{Q_j} +S,
\label{eq:exclusive-process}
\end{equation}
corresponding to the transition $Q_{i} \to Q_{j}+S$. Here and below, $X_{Q}$ denotes a meson which contains a quark $Q$.

The Feynman diagram of the process is shown in Fig.~\ref{fig:Sproduction} (c). Using the Lagrangian~\eqref{eq:g05}, for the matrix element we have
\begin{equation}
    \mathcal{M} (X_{Q_{i}} \to S X'_{Q_{j}}) =
\frac{\theta}{2}\frac{m_{Q_{i}}}{v}\times  \xi_{ij}\times M_{XX'}(m_{S}^{2}),
    \label{eq:matrixelementhhS}
\end{equation} 
where
\begin{equation}
    M_{XX'}((p_{X}-p_{X'})^{2}) \equiv \langle X'(p_{X'})|\bar{Q}_{i}(1+\gamma_{5})Q_{j}|X(p_{X})\rangle
    \label{eq:hadronic-matrix-element-1}
\end{equation}
is the matrix element of the transition $X_{Q_{i}}\to X'_{Q_{j}}$. Expressions for these matrix elements for different initial and final mesons are given in Appendix~\ref{sec:hadronic-form-factors}. So, we can calculate the branching fraction of the corresponding process by the formula ~\cite{Tanabashi:2018oca}
\begin{equation}
    \text{BR}(X_{Q_{i}}\to X'_{Q_{j}}+S) = \frac{1}{\Gamma_{X}}\theta^{2}\frac{|\xi_{ij}|^{2}m_{Q_{i}}^{2}|M_{XX'}(m_{S}^{2})|^{2}}{32\pi v^{2}}\frac{|\bm{p}_{S}|}{m_{X}^{2}},
    \label{eq:productionBRmesons}
\end{equation}
where $\Gamma_{X}$ is the decay width of the meson $X$. We use the lifetimes of mesons from~\cite{Tanabashi:2018oca}.

For the kaons, the only possible 2-body decay is the process
\begin{equation}
    K \to \pi +S
\end{equation}
There are 3 types of the kaons -- $K^{\pm}, K^{0}_{L}, K^{0}_{S}$. Although the decay width for each of them is by given by the same loop factor, $\xi_{sd}$, the branching ratios differ. The first reason is that these kaons have different decay widths. The second reason is that the $K^{0}_{S}$ is approximately the $CP$-even eigenstate. Therefore the decay $K^{0}_{S}\to \pi S$ is proportional to the CKM $CP$-violating phase and is strongly suppressed~\cite{Leutwyler:1989xj}. Further we assume that the corresponding branching ratio vanishes. See Table~\ref{tab:BR} for the branching ratios of $K^{0}_{L},K^{\pm}$.

\subsection{Scalar production in leptonic decays of mesons}
\label{sec:leptonic-decays}
Consider the process $X \to S e \nu$. Its branching ratio is~\cite{Dawson:1989kr, Cheng:1989ib}
\begin{equation}
    \text{BR} (X \rightarrow S e \nu ) =
  \dfrac{\sqrt{2} G_F m_X^4}{96 \pi^2 m_{\mu}^2 (1 - m_{\mu}^2/m_h^2)^2} 
  \times \text{BR} (X \rightarrow \mu \nu) \left(\frac{7}{9}\right)^2 f\left( \frac{m_S^2}{m_X^2}\right),
\end{equation}
where $f(x) = (1 - 8x + x^2) (1 - x^2) - 12 x^2 \text{ln} x$.
The values of the branching ratios for different types of the mesons are shown in Table~\ref{tab:BR-3-body}.
\begin{table}[h]
\centering
\begin{tabular}{|c|c|}
 \hline  
 Meson  & BR$(h \rightarrow S e \nu )/ f\left( x\right) \theta^2$\\
 \hline
 $D \to  S e \nu$ & $5.2 \cdot 10^{-9}$\\
 $K \rightarrow S e \nu$ & $4.1 \cdot 10^{-8}$ \\
 $B \rightarrow S e \nu$ & $< 7.4 \cdot 10^{-10}$  \\
 \hline
\end{tabular}
    \caption{Branching ratios of 3-body meson decay. From experimental data we have only upper bound on the $\text{BR}(B\to\mu\nu)$, so we put upper bound on $B \rightarrow S e \nu$ decay.}
    \label{tab:BR-3-body}
\end{table}
However, although for the $D$ this channel enhances the production in $\simeq \mathcal{O}(100)$ times, the production from $D$ is still sub-dominant.

\section{DIS}
\label{sec:production-direct-dis}
The scalar production in the DIS is driven by the interaction with the quarks and gluons:
\begin{equation}
    \mathcal{L} =  S\theta\sum_{q}\frac{m_{q}}{v}\bar{q}q + \theta\frac{S\alpha_{s}}{16\pi v}F_{G}(m_{S})G_{\mu\nu}^{a}G^{\mu\nu ,a},
    \label{eq:dis-lagrangian}
\end{equation}
where $F_{G}$ is a loop factor being of order of $|F_{G}|^{2} \simeq 10-20$ for the scalars in the mass range $m_{S} \lesssim 10\text{ GeV}$ (see Appendix~\ref{sec:effective-interactions-gauge-bosons}). Processes of the scalar production in DIS are quark and gluon fusions:
\begin{equation}
q+\bar{q}\to S, \quad G + G \to S
\label{eq:dis-processes}
\end{equation}
\begin{figure}[h!]
  \centering
    \includegraphics[width=\textwidth]{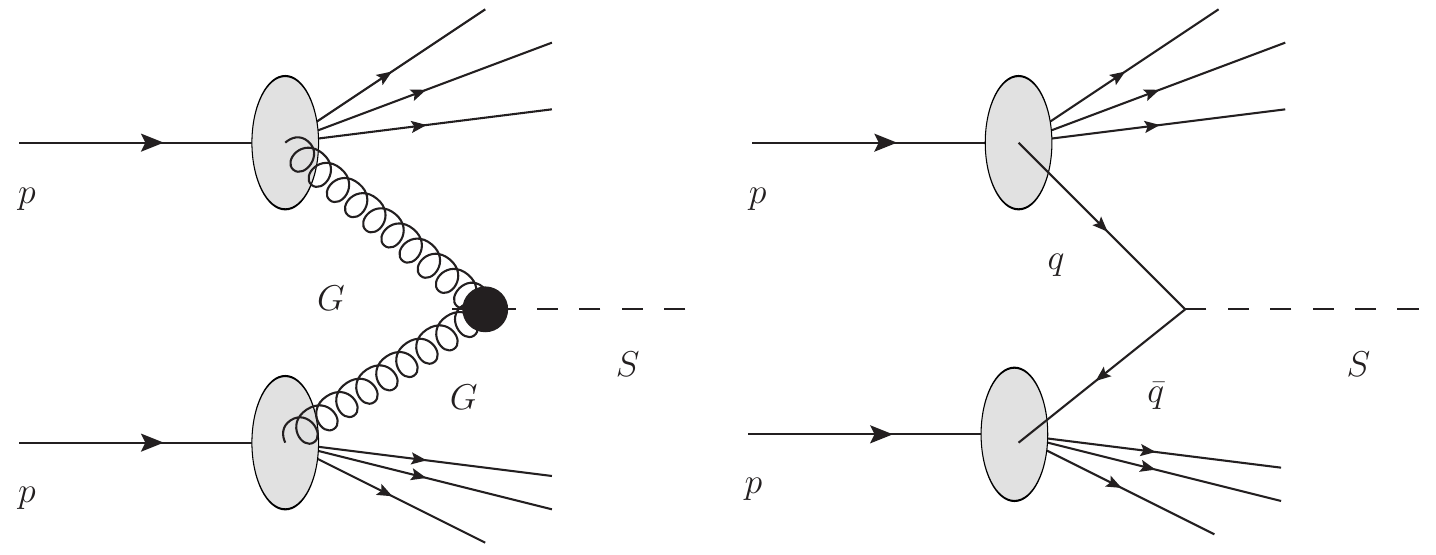}
    \caption{The diagrams of the production of the scalar in deep inelastic scattering.}
    \label{fig:dis-production}
\end{figure}
Corresponding diagrams are shown in Fig.~\ref{fig:dis-production} and the matrix elements are
\begin{equation}
    \mathcal{M}(GG\to S) = 4\frac{F_{G}(m_{S}) \alpha_s}{16 \pi} \frac{\theta}{v} [(k_2^{\mu} \cdot k_1^{\nu}) - g^{\mu \nu} (k_1 \cdot k_2)] \epsilon_{\mu}(k_1) \epsilon_{\nu} (k_2),
\end{equation}

\begin{equation}
    i\mathcal{M}(q\bar{q}\to S) = 
    \overline{v}(k_2) \left( \frac{-i \theta \overline{m}_q}{v} \right) u(k_1).
\end{equation}
The differential cross section is given by
\begin{equation}
    d \sigma(s_{YY}) = 
    \frac{(2 \pi)^4}{4} 
    \frac{\overline{|\mathcal{M}(YY\to S)|^2}}{\sqrt{(k_1 \cdot k_2)^2}}
    d \Phi (k_1 + k_2,p_{S}) = \frac{\pi \overline{|\mathcal{M}(YY\to S)|^2}}{ m_S^2} \delta(s_{YY} - m_S^2),
    \label{eq:differential-cross-section-dis}
\end{equation}
where Y denotes a quark/antiquark or a gluon, $\overline{|\mathcal{M}(GG\to S)|^{2}}$ is the squared matrix element averaged over gluon or quark polarizations and
\begin{equation}
d \Phi(k_1 + k_2,p_{S}) = 
\delta^4(k_1 + k_2 - p_{S}) 
\dfrac{d^3 \bm{p}_{S}}{(2 \pi)^3 2 E_s}.
\end{equation}

The hard cross sections for the gluon and quark fusions are thus
\begin{equation}\label{sigmagg}
    \sigma_{G}(s_{GG})  = 
    \delta(s_{GG} - m_S^2) \frac{|F_{G}(m_{S})|^2 \alpha_s^2 \theta^2 m_S^2}{128\pi v^2},
\end{equation}
\begin{equation}
\label{hard_cross_section_quarks}
    \sigma_{q}(s_{qq}) = 
    \frac{\pi}{m_S^2} 
    \delta(s_{qq} - m_S^2) 
    \frac{\theta^2 \overline{m}_q^2}{2v^2} m_S^2.
\end{equation}
Using hard cross sections \eqref{hard_cross_section_quarks} and \eqref{sigmagg}, one can  calculate the total cross section of the production in DIS as
\begin{equation}
    \sigma_{\text{DIS},Y}=g_{Y}\int \sigma_{Y}(s)f_{Y_{1}}(\sqrt{s_{Y_{1}Y_{2}}},x_{1}) f_{Y_{2}}(\sqrt{s_{Y_{1}Y_{2}}},x_{2}) dx_1 dx_2.
\end{equation}
Here, $f_{Y}(Q,x)$ is the parton distribution function (pdf) of the parton $Y$ carrying the momentum fraction x at the scale $Q$. $g_{q} = 2, g_{G} = 1$; $g_{q}$ is a combinatorial factor taking into account that the quark/antiquark producing a scalar can be stored in both of colliding protons.

The result is
\begin{equation}
    \sigma_{\text{DIS},q} (s) 
    = \frac{\pi}{m_S^2} 
    \frac{\theta^2 \overline{m}_q^2 m_S^2}{2v^2 s}
    \times W_{q\bar{q}}, \ \sigma_{\text{DIS},G} (s) 
    = \theta^{2}\frac{|F_{G}(m_{S})|^{2}\alpha_{s}^{2}(m_{S})m_{S}^{2}}{128\pi sv^{2}}\times W_{GG}.
    \label{eq:dis-cross-sections}
\end{equation}
Here, $s$ denotes the $pp$ CM energy, $\overline{m}_{q}$ is the $\overline{\text{MS}}$ quark mass at the scale $m_{S}$, and
\begin{equation}
 W_{XX}(s,m_{S}) \equiv \int\limits_{m_S^2/s}^1 \frac{dx}{x} f_{X}(m_{S},x) f_{X}\left(m_{S}, \frac{m_S^2}{sx} \right)
 \label{eq:partonic-weight}
\end{equation}
is the partonic weight of the process. Since the partonic model breaks down at scales $Q \lesssim 1\text{ GeV}$, the description of the scalar production in DIS presented in this section is valid only for scalars with masses $m_{S}\gtrsim 1\text{ GeV}$. For numerical estimates we have used LHAPDF package~\cite{Buckley:2014ana} with CT10NLO pdf set~\cite{Lai:2010vv}.

The main contribution to the DIS cross section comes from gluons. To see this, let us compare the gluon cross section $\sigma_{\text{DIS},G}$ with the s-quark cross section $\sigma_{\text{DIS},S}$, which is the largest quark cross section.\footnote{Indeed, the quark cross sections are proportional to the Yukawa constant squared $y_{q}^{2}$, and the large ratio $(y_{s}/y_{u,d})^{2}$ compensates smaller partonic weight $W_{ss}/W_{u,d}$.} Their ratio is
\begin{equation}
    \frac{\sigma_{\text{DIS},G}}{\sigma_{\text{DIS},s}}\approx 0.6\left(\frac{\alpha_{s}(m_{S})}{0.4}\right)^{2}\frac{|F_{G}(m_{S})|^{2}}{20}\times \left(\frac{m_{S}}{1\text{ GeV}}\right)^{2}\frac{W_{GG}}{W_{ss}}.
\label{eq:dis-gluon-to-quark}
\end{equation}
The product $|F_{G}|^{2} \alpha_{s}^{2}$ changes with $m_{S}$ relatively slowly, and therefore the ratio~\eqref{eq:dis-gluon-to-quark} is determined by the product $(m_{S}/1\text{ GeV})^{2}W_{GG}/W_{ss}$. It is larger than one for the masses $m_{S}\gtrsim 2 \text{ GeV}$ in broad CM energy range, see Fig.~\ref{fig:dis-production-plot}.

Having the cross sections~\eqref{eq:dis-cross-sections}, we calculate the DIS probability as
\begin{equation}
    P_{\text{DIS}} = \frac{\sum_{q}\sigma_{\text{DIS},q}+\sigma_{\text{DIS},G}}{\sigma_{pp}},
    \label{eq:dis-probability}
\end{equation}
where for the total proton-proton cross section $\sigma_{pp}$ we used the data from~\cite{Tanabashi:2018oca}.

\section{Scalar production in proton bremsstrahlung}
\label{sec:bremsstrahlung}
\begin{figure}[h!]
    \centering
    \includegraphics[width=0.6\textwidth]{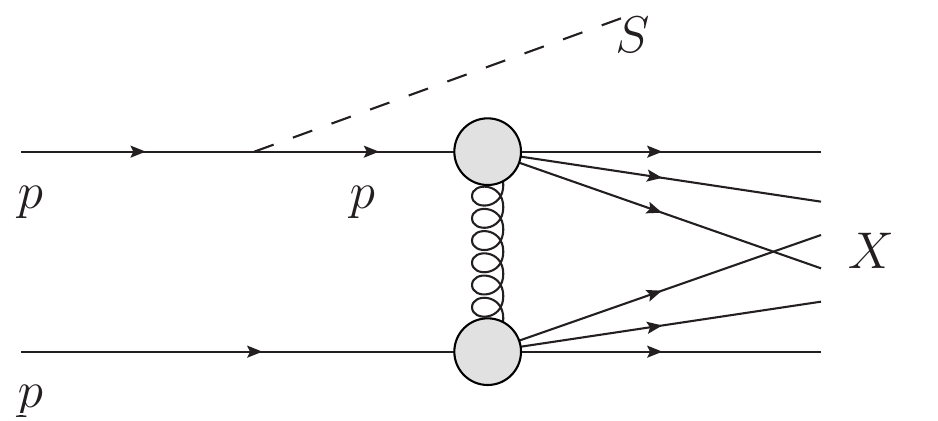}
    \caption{A diagram of the production of a scalar in the proton bremsstrahlung process.}
    \label{fig:bremsstrahlung}
\end{figure}
A scalar $S$ can be produced through the $SNN$ vertex (see Sec.~\ref{sec:effective-interactions-nucleons}) in proton-proton bremsstrahlung process
\begin{equation}
p+p \to S+X,
\label{eq:bremsstrahlung-process}
\end{equation}
with the diagram of the process shown in Fig.~\ref{fig:bremsstrahlung}. Corresponding probability can be estimated using generalized Weizsacker-Williams method, allowing to express the cross section of the given process by the cross section of its sub-process~\cite{Altarelli:1977zs,Blumlein:2013cua,Blumlein:2011mv,Chen:1975sh,Kim:1973he,Frixione:1993yw,Baier:1973ms}. Namely, let us denote the momentum of the incoming proton in the rest frame of the target proton by $p_{p}$, the fraction of $p_{p}$ carried by $S$ as $z$ and the transverse momentum of $S$ as $p_{T}$. Then, under conditions
\begin{equation}
        \frac{p_{T}^{2}}{4p_{p}^{2}}\ll z(1-z)^{2}, \quad \frac{m_{S}^{2}}{4p_{p}^{2}}\ll z(1-z), \quad \frac{m_{p}^{2}}{4p_{p}^{2}}\ll \frac{(1-z)^{2}}{z}
        \label{eq:bremsstrahlung-domain}
        \end{equation}
the differential production cross section of $S$ production can be written as (see Appendix~\ref{sec:bremsstrahlung-derivation})
\begin{equation}
d\sigma_{\text{brem}} \approx \sigma_{pp_{t}}(s')\times P_{p\to pS}(p_{T},z)dp_{T}^{2}dz,
\end{equation}
where we denoted a target proton as $p_{t}$, $\sigma_{pp_{t}}$ is the total $p$-$p$ cross section, $s' = 2m_{p}p_{p}(1-z)+2m_{p}^{2}$ and the differential splitting probability of the proton to emit a scalar is
\begin{equation}
    P_{p\to pS}(p_{T},z) \approx |F_{pS}(m_{S}^{2})|^{2} \frac{g_{SNN}^{2}\theta^{2}}{8\pi^{2}}z\frac{m_{p}^{2}(2-z)^{2}+p_{T}^{2}}{(m_{p}^{2}z^{2}+m_{S}^{2}(1-z)+p_{T}^{2})^{2}},
\end{equation}
with $g_{SNN}$ being low-energy proton-scalar coupling, and $F_{pS}$ the scalar-proton form-factor, see Appendix~\ref{sec:effective-interactions-nucleons} and Eq.~\eqref{eq:scalar-nucleon-form-factor}. 

\begin{figure}[h!]
    \centering
    \includegraphics[width=0.5\textwidth]{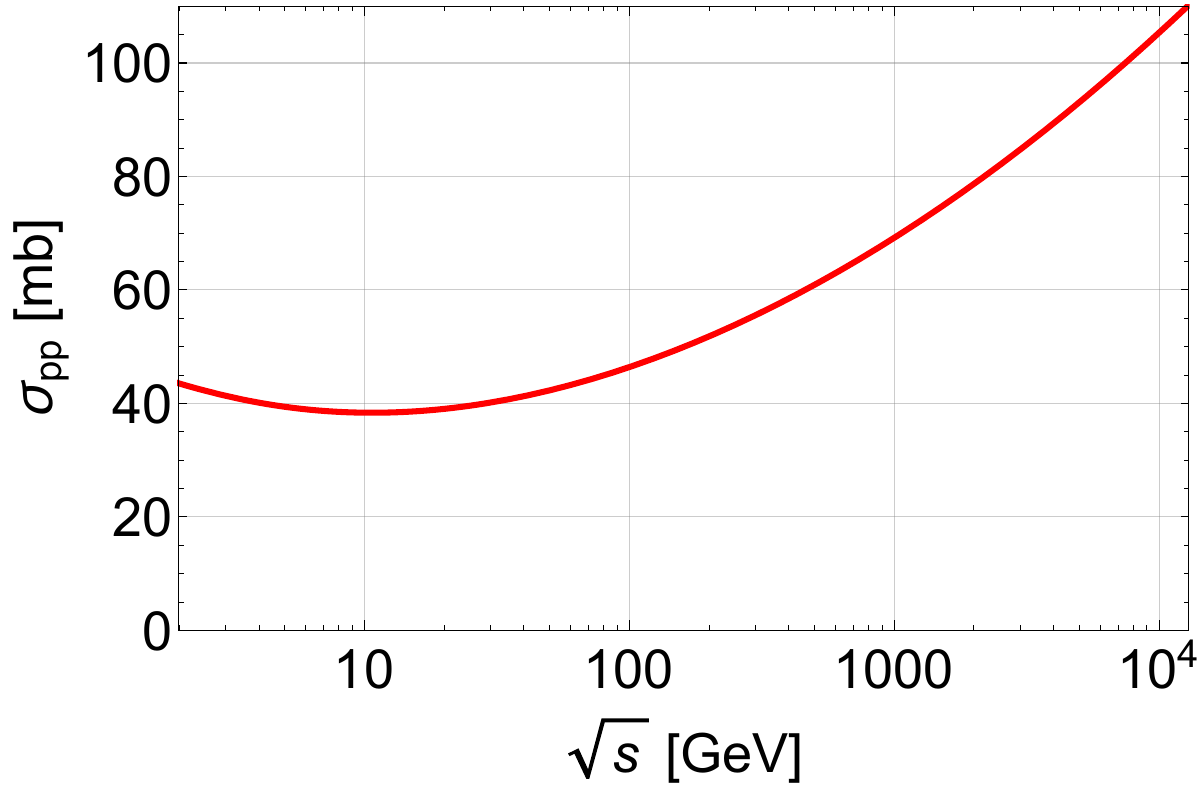}~\includegraphics[width=0.5\textwidth]{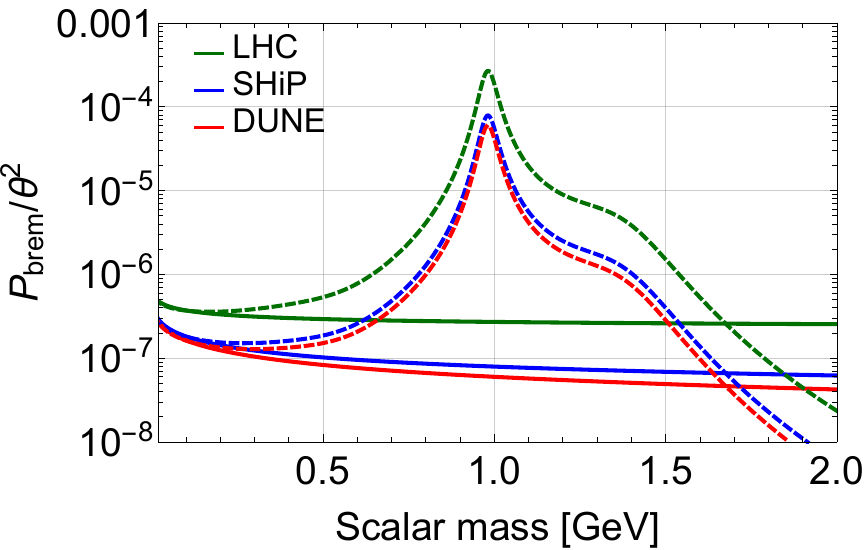}
    \caption{\textit{Left panel}: proton-proton total cross section as a function of the center of mass energy $\sqrt{s_{pp}}$. \textit{Right panel}: the probability of the production of a scalar $S$ in bremsstrahlung process versus the scalar mass. The solid line assumes the unit form-factor $F_{pS}(q^{2}) = 1$, while the dashed line corresponds to the form-factor from Eq.~\eqref{eq:scalar-nucleon-form-factor}.}
    \label{fig:pp-cross-section}
\end{figure}

For the total $pp$ cross section, we use the experimental fit
\begin{equation}
    \sigma_{pp}(s) = Z+B \ln^{2}\left(\frac{s}{s_{0}}\right)+C_{1}\left( \frac{s_{1}}{s}\right)^{\eta_{1}}-C_{2}\left( \frac{s_{1}}{s}\right)^{\eta_{2}},
\end{equation}
where $Z = 35.45\text{ mb}$, $B = 0.308\text{ mb}$, $C_{1}=42.53\text{ mb}$, $C_{2} = 33.34\text{ mb}$, $\sqrt{s_{0}}=5.38\text{ GeV}$, $\sqrt{s_{1}} = 1\text{ GeV}$, $\eta_{1}=0.458$ and $\eta_{2} = 0.545$~\cite{Tanabashi:2018oca}. This cross section is shown in Fig.~\ref{fig:pp-cross-section}, from we see that it is almost constant for a wide range of energies.

The total cross section can be written in the form
\begin{equation}
    \sigma_{\text{brem}} = g_{SNN}^{2}\theta^{2}|F_{pS}(m_{S}^{2})|^{2}\sigma_{pp}(s)\mathcal{P}_{\text{brem}}(s,m_{S}),
\end{equation}
where 
\begin{equation}
    \mathcal{P}_{\text{brem}}(s,m_{S}) = \frac{1}{g_{SNN}^{2}\theta^{2}}\int dp_{T}^{2}dz P_{p\to pS}(p_{T},z)\frac{\sigma_{pp}(s')}{\sigma_{pp}(s)}.
\end{equation}
The domain of the definition of $p_{T}$ and $z$ is determined by the conditions~\eqref{eq:bremsstrahlung-domain}. For definiteness, we fix the domain of integration by the requirement
\begin{equation}
    \frac{m_{S}^{2}(1-z)+m_{p}^{2}z^{2}+p_{T}^{2}}{4p_{p}^{2}z(1-z)^{2}}<0.1.
\end{equation}
The probability of a scalar production in proton bremsstrahlung is
\begin{equation}
    P_{\text{brem}} = \frac{\sigma_{\text{brem}}}{\sigma_{pp}(s)} \approx g_{SNN}^{2}\theta^{2}|F_{pS}(m_{S}^{2})|^{2}\mathcal{P}_{\text{brem}}(s,m_{S})  ,
\end{equation}
where $s$ is the CM energy of two protons. We show its dependence on the scalar mass and the incoming beam energy in Fig.~\ref{fig:pp-cross-section}.

\subsection{Splitting probability derivation}
\label{sec:bremsstrahlung-derivation}
\begin{figure}[h!]
    \centering
    \includegraphics{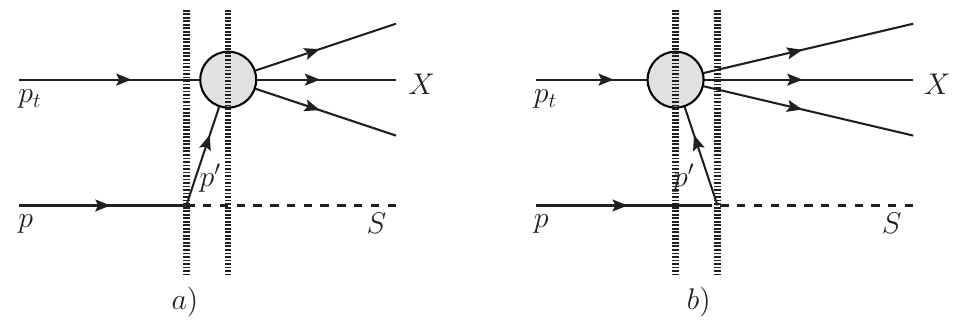}
    \caption{The lowest order old-fashioned perturbation theory diagrams for the bremsstrahlung process~\eqref{eq:bremsstrahlung-process}. Vertical dotted lines denote the intermediate states.}
    \label{fig:bremsstrahlung-old-fashioned}
\end{figure}

Following the approach described in~\cite{Altarelli:1977zs}, let us consider the process~\eqref{eq:bremsstrahlung-process} within the old-fashioned perturbation theory. The corresponding diagrams are shown in Fig.~\ref{fig:bremsstrahlung-old-fashioned}.
The matrix element has the form $\mathcal{V}_{pp_{t}\to SX} = \mathcal{V}_{a}+\mathcal{V}_{b}$, where
\begin{equation}
    \mathcal{V}_{a} = \frac{\mathcal{M}_{p\to p'S}\mathcal{M}_{p'p_{t}\to X}}{2E_{p'}(E_{p'}+E_{S}-E_{p})}\bigg|_{\bm{p}_{p'} = \bm{p}_{p}-\bm{p}_{S}}, \quad \mathcal{V}_{b} = \frac{\mathcal{M}_{p_{t}\to p'X}\mathcal{M}_{p'p\to S}}{2E_{p'}(E_{p}+E_{p'}-E_{S})}\bigg|_{\bm{p}_{p'} = \bm{p}_{p}-\bm{p}_{S}}.
    \label{eq:matrix-element-old-fashioned}
\end{equation}
Here, $\mathcal{M}$ denotes Lorentz-invariant amplitude of the processes. There exists a kinematic domain at which $|\bm{M}_{b}| \ll |\bm{M}_{a}|$. Namely, let us consider an ultrarelativistic incoming $p$, and write the 4-momenta of $p$, $S$ and intermediate $p'$ as
\begin{align}
P_{p}^{\mu} &= \left(p_{p}+ \frac{m_{p}^{2}}{2p_{p}^{2}}, \bm{0}, p_{p}\right), \\  
P_{S}^{\mu} &= \left(p_{p}z +\frac{p_{T}^{2}+m_{S}^{2}}{2p_{p}z},\bm{p}_{T},zp_{p}\right), \\ 
P_{p'}^{\mu} &= \left((1-z)p_{p}+\frac{m_{p}^{2}+p_{T}^{2}}{2p_{p}(1-z)},-\bm{p}_{T},(1-z)p_{p} \right),   
\end{align}
where $p_{T}$ is a transverse momentum of $S$ and $z$ is a fraction of a parallel momentum carried by $S$. Then the energy denominators in~\eqref{eq:matrix-element-old-fashioned} are
\begin{equation}
   \Delta E_{a} = E_{p'}+E_{S}-E_{p} \approx \frac{p_{T}^{2}+(1-z)m_{S}^{2}+z^{2}m_{p}^{2}}{2p_{p}z(1-z)}, \quad \Delta E_{b} = E_{p}+E_{p'}-E_{S} \approx 2p_{p}(1-z).
\end{equation}
Assuming that $\Delta E_{a}\ll \Delta E_{b}$ we can neglect the matrix element $\mathcal{V}_{b}$. 

\begin{figure}[h!]
    \centering
    \includegraphics[width=\textwidth]{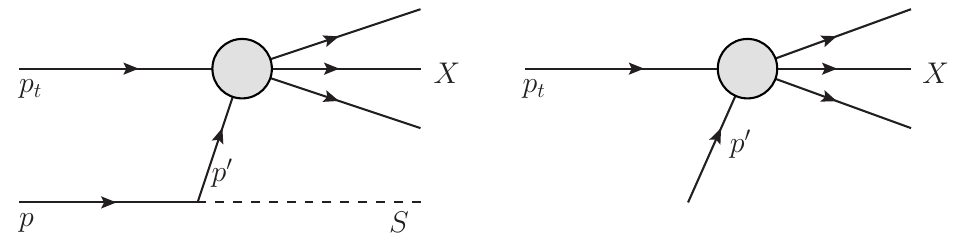}
    \caption{Diagrams the bremsstrahlung process~\eqref{eq:bremsstrahlung-process}  (left) and its sub-process $pp_{t} \to X$ describing a proton-proton collision (right).}
    \label{fig:bremsstrahlung-process-sub-process}
\end{figure}

Once we neglect $\mathcal{V}_{b}$, it is possible to relate the differential cross section of the process~\eqref{eq:bremsstrahlung-process} to the total $pp$ scattering cross section. Indeed, let us consider a corresponding process $pp_{t}\to X$, which is a sub-process of~\eqref{eq:bremsstrahlung-process} obtained by removing the in $p$ line and out $S$ line, see Fig.~\ref{fig:bremsstrahlung-process-sub-process}.
The matrix element for this process is simply
\begin{equation}
\mathcal{V}_{pp_{t}\to S} = \mathcal{M}_{pp_{t}\to X}.
\label{eq:matrix-element-pp-inelastic}
\end{equation}
Using~\eqref{eq:matrix-element-old-fashioned},~\eqref{eq:matrix-element-pp-inelastic}, for the corresponding differential cross sections we obtain
\begin{multline}
    d\sigma_{pp_{t}\to SX} = \frac{1}{4E_{p}E_{p_{t}}}\frac{|\mathcal{M}_{p\to p'S}|^{2}|\mathcal{M}_{p'p_{t}\to X}|^{2}}{(2E_{p'})^{2}(E_{p'}+E_{S}-E_{p})^{2}}\times \\ (2\pi)^{4}\delta^{(4)}\left(p_{p}+p_{p_{t}}-p_{S}-\sum_{X}p_{X}\right) \frac{d^{3}\bm{p}_{S}}{(2\pi)^{3}2E_{S}}\times \prod_{X}\frac{d^{3}\bm{p}_{X}}{(2\pi)^{3}2E_{X}},
    \label{eq:sub-process-a}
\end{multline}
\begin{equation}
    d\sigma_{p'p_{t}\to X} = \frac{1}{4E_{p'}E_{p_{t}}}|\mathcal{M}_{p'p_{t}\to X}|^{2}\times (2\pi)^{4}\delta^{(4)}\left(p_{p'}+p_{p_{t}}-\sum_{X}p_{X}\right) \prod_{X}\frac{d^{3}\bm{p}_{X}}{(2\pi)^{3}2E_{X}}
    \label{eq:sub-process-b}
\end{equation}
Neglecting the difference in the energy conservation arguments in the delta-functions that are of order $\mathcal{O}(m_{p/S}^{2}/p_{p}^{2},p_{T}^{2}/p_{p}^{2})$, we can relate these two cross sections as
\begin{equation}
    d\sigma_{pp_{t}\to SX} = dP_{p\to p'S}(z,p_{T})d\sigma_{p'p_{t}\to X}(p_{T},z),
    \label{eq:differential-cross-section}
\end{equation}
where we introduced differential splitting probability $dP_{p\to p'S}$: 
\begin{equation}
 dP_{p\to p'S}(p_{T},z) \equiv 2\frac{|\mathcal{M}_{p\to p'S}|^{2}}{4E_{p}E_{p'}(E_{p'}+E_{S}-E_{p})^{2}}\frac{d^{3}\bm{p}_{S}}{(2\pi)^{3}2E_{S}}.
    \label{eq:splitting-probability}
\end{equation}
Here a factor of $2$ is combinatorial factor taking into account that a scalar can be produced from both the legs of colliding protons.

Integrating the differential cross section~\eqref{eq:differential-cross-section} over the momenta of the final states particles $X$ and summing over all possible sets $\{X\}$, we finally arrive at
\begin{equation}
    d\sigma_{pp_{t}\to SX} \approx P_{p\to p'S}(z,p_{T})dp_{T}^{2}dz\sigma_{pp}(s'),
\end{equation}
where $s' \approx 2m_{p}p_{p}(1-z)+2m_{p}^{2}$~\footnote{Here we neglected the $p_{T}$ dependence in $\sigma_{pp}$.} and $\sigma_{pp}(s')$ is the total proton-proton cross section.

Let us now find explicit expression for the splitting probability~\eqref{eq:splitting-probability}. Using the expressions~\eqref{eq:bremsstrahlung-domain}, we find 
\begin{equation}
\frac{d^{3}\bm{p}_{S}}{(2\pi)^{3}2E_{S}} \approx \frac{dp_{T}^{2}dz}{16 \pi^{2}z}, \quad 
|\mathcal{M}_{p\to pS}|^{2} \approx 2g_{SNN}^{2}\theta^{2}|F_{pS}(m_{S}^{2})|^{2}(m_{p}^{2}+(P_{p}\cdot P_{p'})).
\end{equation}
Finally, we arrive at
\begin{equation}
P_{p\to p'S} \approx \frac{g_{SNN}^{2}\theta^{2}|F_{pS}(m_{S}^{2})|^{2}}{8\pi^{2}}z\frac{m_{p}^{2}(2-z)^{2}+p_{T}^{2}}{(m_{p}^{2}z^{2}+m_{S}^{2}(1-z)+p_{T}^{2})^{2}}.
\end{equation}

\section{Scalar production in photon fusion}
\label{sec:production-direct-coherent}
\begin{figure}[h!]
    \centering
    \includegraphics[width=0.4\textwidth]{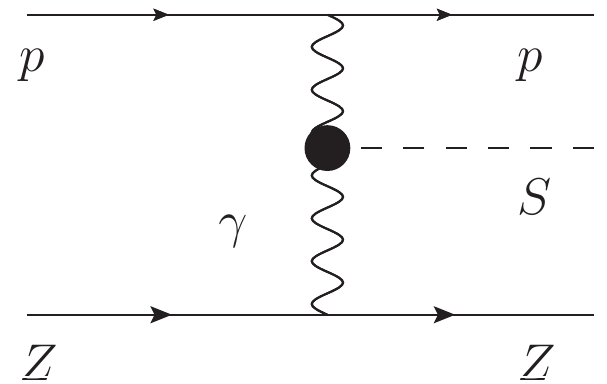}
    \caption{A diagram of the production of a scalar in photon fusion.}
    \label{fig:photon-fusion}
\end{figure}

\begin{figure}[h!]
    \begin{center}
        {
      \includegraphics[width=0.7\textwidth]{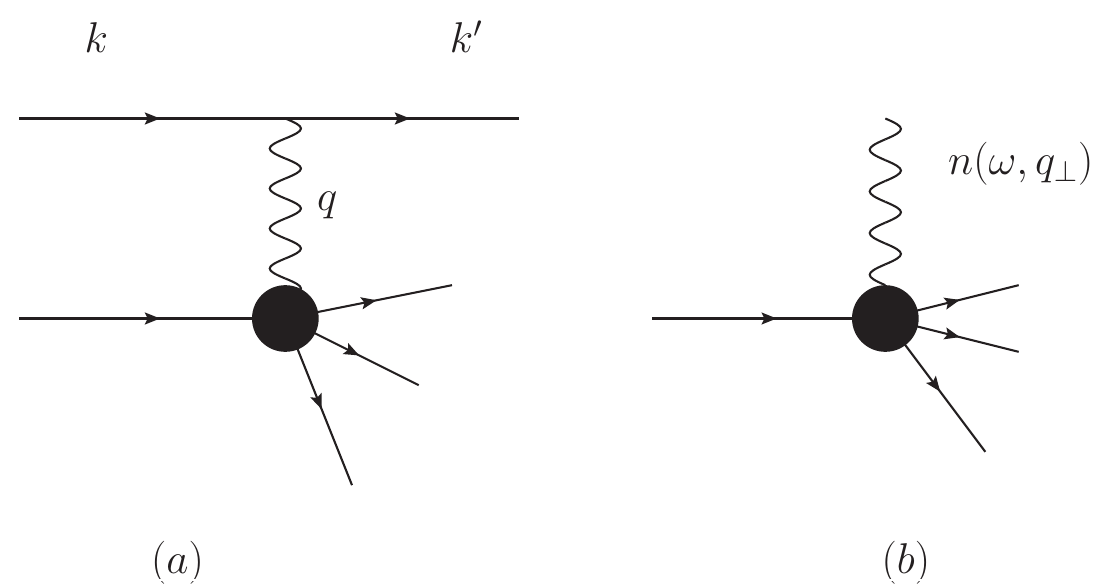}
        }
    \end{center}
    \caption{The idea of the equivalent photon approximation. If a charge with the momentum $k$, emitting the virtual photon with the virtuality $q$, is ultrarelativistic, then the cross section of the process (a) can be expressed in terms of the cross section of the process (b). The remained effect of the charge is the distribution function $n_{\text{charge}}(x,q^{2})$, where $x$ is the energy fraction carried by photon.}
    \label{fig:EPA}
\end{figure}

A scalar can be produced elastically in $pp$ collisions through the $S\gamma\gamma$ vertex (see Appendix~\ref{sec:effective-interactions-gauge-bosons}). The production process is
\begin{equation}
    p+Z \to p + Z +S,
    \label{eq:photon-fusion}
\end{equation}
with the corresponding diagram shown in Fig.~\ref{fig:photon-fusion}.
To find the number of produced scalars in the photon fusion, we will use the equivalent photon approximation (EPA), which provides a convenient framework for studying processes involving photons emitted from fast-moving charges~\cite{Budnev:1974de,Martin:2014nqa,Dobrich:2015jyk}. The basic idea of the EPA is a replacement of the charged particle $Y$ in the initial and final state, that interacts through the virtual photon carrying the virtuality $q$ and the fraction of charge's energy $x$, by the almost real photon with a distribution $n_{Y}(x; q^2)$ that depends on the type of the charged particle, see Fig.~\ref{fig:EPA}.
The magnitude of the momentum transfer carried by the virtual photon can be approximated as
\begin{equation}
\label{eq:phf1}
    q^{2}\approx \frac{q_t^2+x^2 m_{Y}^2}{1-x},
\end{equation}
where $q_t$ is the transverse component of the spatial momentum of the photon with respect to the spatial momentum of the particle $Y$, and $m_{Y}$ is the mass of $Y$. Conditions for validity of the EPA are $x\ll 1$ and $q_t\lesssim x m_{Y}$~\cite{Dobrich:2015jyk}. The distribution $n_{Y}(x,q_{t}^{2})$ of the emitted photons can be described by
\begin{equation}
\label{eq:phf3}
n_Y(x; q_t^2)=\frac{\alpha_{\text{EM}}}{2\pi}
\frac{1 + (1 -x)^2}{x(q_t^2+x^2 m_{X}^2)}\left[
\frac{q_t^2}{q_t^2+x^2 m_Y^2}D_{Y}(q^2)+\frac{x^2}2 C_{Y}(q^2)\right],
\end{equation}
where $C(q^{2}),D(y^{2})$ are appropriate form-factors. We take the proton and nucleus form-factors from~\cite{Dobrich:2015jyk}.

Within the EPA, we approximate the cross section of the process~\eqref{eq:photon-fusion} by
\begin{equation}
\label{eq:phf9}
\sigma_{pZ\rightarrow SpZ}=\int dx_1 dx_2 d\vec q^{\,\,2}_{1t} d\vec q_{2t}^{\,\,2}
\gamma_p(q_{1t}^2,x_1)\gamma_Z(q_{2t}^2,x_2)\sigma_{\gamma\gamma\to S}(s_{\gamma\gamma}).
\end{equation}
Here
\begin{equation}
\label{eq:sigmaphiphi}
    \sigma_{\gamma\gamma\to S}(s_{\gamma\gamma}) = 
    \frac{\pi}{ m_S^2} \frac{|F_\gamma(m_S)|^{2} \alpha_{\text{EM}}^2 \theta^2 m_S^4}{256 \pi^2 v^2}\,\delta(s_{\gamma\gamma} - m_S^2) \equiv\frac{1}{x_{1}}\Sigma_{\gamma\gamma} \frac{\delta\left(x_2 - \frac{m_S^2}{x_1 s_{p}Z}\right)}{x_{1}s_{pY}},
\end{equation}
where $s_{\gamma\gamma}=(q_1+q_2)^2 \approx 4x_1x_2E_{p}^{\text{CM}}E_{Y}^{\text{CM}} \approx x_{1}x_{2}s_{pY}$, and 
\begin{equation}
    \Sigma_{\gamma\gamma} = \theta^{2}\frac{|F_{\gamma}|^{2}\alpha_{\text{EM}}^{2}m_{S}^{2}}{256 \pi v^{2}s_{pZ}}.
\end{equation}
Let us discuss the boundaries of integration in Eq.~\eqref{eq:phf9}. Following~\cite{Dobrich:2015jyk}, for the upper limit of $q$ we choose $q_{\text{max}} = 1\text{ GeV}$ for the maximal virtuality of a photon emitter by the proton and $q_{\text{max}} = 4.49/R_{1}$ for a photon emitted by the nucleus. Using~\eqref{eq:phf1}, we get $x_{p,max}\approx 0.63$,  $x_{Z,max}=0.018$. The lower bound on $q$, it is given by the kinematic threshold for the $S$ particle production. For the nucleus, there is additional constraint $q^{2} \gtrsim r_{\text{s}}^{-2}$, where $r_{s} \simeq 10\text{ keV}$ is the inverse radius of the electron shell (at larger scales the nucleus is screened by electrons).

Substituting the photon fusion cross section~\eqref{eq:sigmaphiphi} into~\eqref{eq:phf9}, for the $pZ$ cross section we get
\begin{equation}
    \sigma_{pZ\rightarrow pZS}=Z^{2}\alpha_{\text{EM}}^{2}\Sigma_{\gamma\gamma} \times W_{\text{coh}},
    \label{eq:coherent-cross-section}
\end{equation}
where
\begin{equation}
    W_{\gamma\text{ fusion}} = \frac{(2\pi)^2}{Z^{2}\alpha_{\text{EM}}^{2}}\int\limits_0^{q_{1t,\text{max}}} q_{1t}\, dq_{1t}
    \int\limits_0^{q_{2t,\text{max}}} q_{2t}\, dq_{2t}\int\limits_{\frac{m_S^2}{x_{2,max} s_{pZ} }}^{x_{1,max}} \frac{dx_1}{x_1}\,
    \gamma_p(x_1,q_{1t})\gamma_{Z}\left(\frac{m_S^2}{x_{1}s_{pZ}},q_{2t}\right)
    \label{eq:coherent-weight}
\end{equation}
is the integrated form-factor. Here we simplified the integration domain for $p_{t}$ assuming $q_{1t,\text{max}},q_{2t,\text{max}} = 1\text{ GeV}$, since the integrand is nonzero only in some region of parameters within the integration area, and therefore by increasing of integration limits we will not affect the result. 

The production probability is calculated using the cross section~\eqref{eq:coherent-cross-section} as
\begin{equation}
    P_{\gamma\text{ fusion}} = \frac{\sigma_{pZ\to pZS}}{\sigma_{pZ}},
\end{equation}
where $\sigma_{pZ} \approx 53\ A^{0.77}\text{ mb}$ is the total $pZ$ cross section, with $A$ being the mass number of the nucleus target~\cite{Carvalho:2003pza}. 

\begin{figure}
    \centering
    \includegraphics[width=0.7\textwidth]{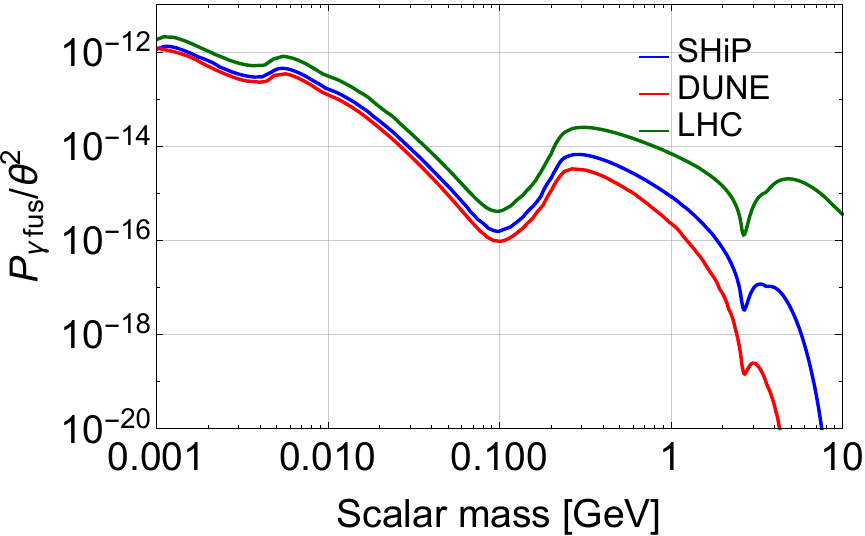}
    \caption{The production probabilities of the scalar in photon fusion process versus the scalar mass. We consider $\text{Mo}$ nucleus ($Z = 42$, $A = 96$).}
    \label{fig:gamma-fusion-probability}
\end{figure}

The dependence of $P_{\gamma \text{ fusion}}$ on the scalar mass and collision energy is shown in Fig.~\ref{fig:gamma-fusion-probability}

\section{Form-factors for the flavor changing neutral current meson decays}
\label{sec:hadronic-form-factors}
Consider matrix elements
\begin{equation}
    M_{XX'}^{P=P'} =\langle X'(p_{X'})|\bar{Q}_{j}Q_{i}|X(p_{X})\rangle, \quad M_{XX'}^{P \neq P'}= \langle X'(p_{X'})|\bar{Q}_{j}\gamma_{5}Q_{i}|X(p_{X})\rangle
    \label{eq:meson-decay-scalar-matrix-element}
\end{equation}
describing transitions of mesons $X(Q_{i}) \to X'(Q_{j})$ in the case of the same and opposite parities $P$, $P'$ correspondingly. These matrix elements can be related to the matrix elements
\begin{equation}
    M^{\mu}_{XX'} \equiv \langle X'(p_{X'})|\bar{Q}_{i}\gamma^{\mu}Q_{j}|X(p_{X})\rangle, \quad M^{\mu 5}_{XX'} \equiv \langle X'(p_{X'})|\bar{Q}_{i}\gamma^{\mu}\gamma_{5}Q_{j}|X(p_{X})\rangle
    \label{eq:meson-decay-vector-matrix-element}
\end{equation}
describing the weak charged current mediating mesons transition $X \to X'$. To derive the relation, we follow~\cite{Bobeth:2001sq} in which a relation for pseudoscalar transition $X'$ was obtained. We generalize this approach to the arbitrary final-state meson. We first notice that
\begin{align}
    M_{XX'}^{P = P'} \equiv \frac{1}{m_{Q_{i}} - M_{Q_{j}}}\langle X'(p_{X'})|\bar{Q}_{i}\slashed{p}_{Q}Q_{j}|X(p_{X})\rangle, \\ M_{XX'}^{P \neq P'} \equiv \frac{1}{m_{Q_{i}} + m_{Q_{j}}}\langle X'(p_{X'})|\bar{Q}_{i}\gamma_{5}\slashed{p}_{Q}Q_{j}|X(p_{X})\rangle,
    \label{eq:meson-decay-scalar-pseudoscalar-matrix-element}
\end{align}
where $p_{Q}^{\mu} \equiv p^{\mu}_{Q_{i}} - p^{\mu}_{Q_{j}}$ and we used the Dirac equation for free quarks. Using then the identity
\begin{equation}
    \bar{Q}_{i}\slashed{p}_{Q}Q_{j} \equiv \hat{P}_{\mu}\bar{Q}_{i}\gamma^{\mu}Q_{j} \equiv [\hat{P}_{\mu},\bar{Q}_{i}\gamma^{\mu}Q_{j}],
\end{equation}
where $\hat{P}_{\mu} \equiv i\partial_{\mu}$ is the momentum operator, we find
\begin{multline}
    M_{XX'}^{P = P'} = \frac{1}{m_{Q_{i}} - m_{Q_{j}}}\langle X'(p_{X'})|[\hat{P}_{\mu},\ \bar{Q}_{i}\gamma^{\mu}Q_{j}]|X(p_{X})\rangle = \\ = -\frac{1}{m_{Q_{i}} - m_{Q_{j}}}(p_{X}-p_{X'})_{\mu}\langle X'(p_{X'})| \bar{Q}_{i}\gamma^{\mu}Q_{j}|X(p_{X})\rangle \equiv -\frac{1}{m_{Q_{i}} - m_{Q_{j}}}q^{\mu}M_{\mu},
    \label{eq:trick}
\end{multline}
where $q_{\mu} \equiv p_{X'\mu} - p_{X\mu}$; for deriving the expression we have acted by $\hat{P}_{\mu}$ on the meson states $|X\rangle, |X'\rangle$. Similarly, for $P\neq P'$ we find
\begin{equation}
     M_{XX'}^{P \neq P'} = -\frac{1}{m_{Q_{i}} + m_{Q_{j}}}q^{\mu}M_{XX\mu}^{5}
\end{equation}
Further we will assume that $X$ is a pseudoscalar, and therefore transitions in pseudoscalar, pseudovector mesons $X'$ are parity even, while transitions in scalar, vector and tensor mesons are parity odd.

\subsection{Scalar and pseudoscalar final meson state}
\subsubsection{Pseudoscalar}
\label{app:pseudoscalar}
In the case of the pseudoscalar meson, $X' = P$, we have~\cite{Ebert:1997mg}
\begin{multline}
    M^{\mu}_{XP} = \langle P(p_{P})|\bar{Q}_{i}\gamma^{\mu}Q_{j}|X(p_{X})\rangle = \\ =\left[(p_{X}+p_{P})^{\mu} - \frac{m_{X}^{2} - m_{P}^{2}}{q^{2}}q^{\mu}\right]f^{XP}_{1}(q^{2})+\frac{m_{X}^{2} - m_{P}^{2}}{q^{2}}q^{\mu}f^{XP}_{0}(q^{2}),
\end{multline}
where $q = p_{X} - p_{P}$.

Contracting it with $q_{\mu}$, we obtain
\begin{equation}
    q_{\mu}M^{\mu}_{XP} = (m_{X}^{2} - m_{P}^{2})f^{XP}_{0}(q^{2})
\end{equation}
Therefore
\begin{equation}
    M_{XP} = \frac{m_{X}^{2} - m_{P}^{2}}{m_{Q_{j}} - m_{Q_{i}}}f^{XP}_{0}(q^{2})
\end{equation}
We take the expression for the form-factor $f^{XP}_{0}(q^{2})$ from~\cite{Ball:2004ye}:
\begin{equation}
\label{eq:pseudoscalar-form-factor}
f^{XP}_{0}(q^{2})=\frac{F_{0}^{XP}}{1-q^{2}/(m^{X}_{\text{fit}})^{2}}
\end{equation}
The values of the parameters $m_{\text{fit}}^{X}$, $F_{0}^{XP}$ for different $X,P$ are summarized in Table~\ref{tab:pseudoscalar-form-factor-parameters}.
\begin{table}[t!]
    \centering
    \begin{tabular}{|c|c|c|c|c|}
        \hline  
        $X, P$ & $B^{+/0}, K^{+/0}$ & $B^{+/0},\pi^{+/0}$ &$K,\pi$  \\
        \hline
        $m_{\text{fit}}^{X},\text{ GeV}$ & $6.16$ & $6.16$ & $\infty$ \\
        \hline
        $F_{0}^{XP}$ & $0.33\pm 0.04$ & $0.258\pm 0.031$ & $0.96$\\
        \hline
    \end{tabular}
    \caption{Values of the parameters in the form-factor~\eqref{eq:pseudoscalar-form-factor} for different $X,P$. We use~\cite{Ball:2004ye,Marciano:1996wy}.}
    \label{tab:pseudoscalar-form-factor-parameters}
\end{table}

\subsubsection{Scalar}
\label{app:scalar}
For the scalar meson $X' = \tilde{S}$ we have~\cite{Sun:2010nv}
\begin{equation}
    M^{\mu}_{X\tilde{S}} = -i\left[(p_{X}+p_{\tilde{S}})^{\mu} -q^{\mu}\right]f^{X\tilde{S}}_{+}(q^{2})
\end{equation}
(here we used $f_{+}(q^{2}) = -f_{-}(q^{2})$ in Eq.~(6) of~\cite{Sun:2010nv}). Similarly to the case $h' = P$,
\begin{equation}
    M_{X\tilde{S}} =  i\frac{m_{X}^{2} - m_{\tilde{S}}^{2} - q^2}{m_{Q_{j}} + m_{Q_{i}}}f^{X\tilde{S}}_{+}(q^{2}).
\end{equation}
Consider the transition $B\to K_{0}^{*}S$. There is an open question whether hypothetical $K_{0}^{*}(700)$ is a state formed by two or four quarks, see, e.g.~\cite{Daldrop:2012sr},  discussions in~\cite{Sun:2010nv,Cheng:2013fba} and references therein. We assume that $K_{0}^{*}(700)$ is a di-quark state and $K_{0}^{*}(1430)$ is its excited state. There are no experimentally observed decays $B \to K_{0}^{*}(700)X$, and therefore there is quite large theoretical uncertainty in determination of the form-factors (see a discussion in~\cite{Issadykov:2015iba}). We will use~\cite{Sun:2010nv}, where there are results for $B\to K_{0}^{*}(700)$ and $B\to K_{0}^{*}(1430)$, and the results for the latter are in good agreement with the experimental data for $B \to K_{0}^{*}(1430)\eta'$ decay. 

We fit the $q^{2}$ dependence of $f^{BK_{0}^{*}}_{+}$ from~\cite{Sun:2010nv} by the standard pole-like function that is used in the literature discussing the $B \to K^{*}_{0}$ transitions (see, e.g.,~\cite{Cheng:2013fba}):
\begin{equation}
    f^{BK_{0}^{*}}_{+}(q^{2}) = \frac{F^{BK_{0}^{*}}_{0}}{1-a\frac{q^{2}}{m_{B}^{2}}+b\left(\frac{q^{2}}{m_{B}^{2}}\right)^{2}},
    \label{eq:scalar-form-factor}
\end{equation}
where $m_{B} = 5.3\text{ GeV}$ is the mass of the $B^{+}$ meson.
The fit parameters are given in Table~\ref{tab:scalar-form-factor-parameters}.
\begin{table}[t!]
    \centering
    \begin{tabular}{|c|c|c|c|c|}
        \hline  
        $\tilde{S}$ &$F_{0}^{B\tilde{S}}$ &    $a$ & $b$   \\
        \hline
         $K_{0}^{*}(700)$ & $0.46$ & $1.6$ & $1.35$ \\
        \hline
        $K_{0}^{*}(1430)$ & $0.17$  & $4.4$  & $6.4$ \\
        \hline
    \end{tabular}
    \caption{Values of the parameters in the form-factor~\eqref{eq:scalar-form-factor} for $B = B^{+}$, $\tilde{S} = K_{0}^{*0}(700)$, $K_{0}^{*}(1430)$. We used~\cite{Sun:2010nv}.}
    \label{tab:scalar-form-factor-parameters}
\end{table}

\subsection{Vector and pseudovector final meson state}
\subsubsection{Vector}
\label{app:vector}
For the vector final state, $X' = V$, we have~\cite{Ebert:1997mg,Ball:2004rg}
\begin{multline}
    \langle V(p_{V})| \bar{Q}_{i}\gamma^{\mu}\gamma_{5}Q_{j} |X(p_{X})\rangle = (m_{X}+m_{V})\epsilon^{\mu *}(p_{V})A_{1}(q^{2}) - \\ -(\epsilon^{*}(p_{V})\cdot q)(p_{X} + p_{V})^{\mu}\frac{A_{2}(q^{2})}{m_{X}+m_{V}}-2m_{V}\frac{\epsilon^{*}(p_{V})\cdot q}{q^{2}}q^{\mu}(A_{3}(q^{2})-A_{0}(q^{2})),
    \label{eq:vector-form-factor-axial-part}
\end{multline}
\begin{equation}
     \langle V(p_{V})| \bar{Q}_{i}\gamma^{\mu}Q_{j} |X(p_{X})\rangle= \frac{2V(q^{2})}{m_{X}+m_{V}}i\epsilon^{\mu\nu\rho\sigma}\epsilon^{*}_{\nu}(p_{V})p_{X, \rho}p_{V, \sigma},
     \label{eq:vector-form-factor-vector-part}
\end{equation}
where $\epsilon_{\mu}(p_{V})$ is the polarization vector of the vector meson, and $A_{i}, V$ are the form-factors. The form-factor $A_{3}$ is related to $A_{1}$ and $A_{2}$ as
\begin{equation}
    A_{3}(q^{2}) = \frac{m_{X}+m_{V}}{2m_{V}}A_{1}(q^{2})-\frac{m_{X} - m_{V}}{2m_{V}}A_{2}(q^{2})
    \label{eq:a3-a0}
\end{equation}
Contracting~\eqref{eq:vector-form-factor-axial-part} and~\eqref{eq:vector-form-factor-vector-part} with $q_{\mu}$, we obtain that the vector part of the matrix element vanishes, while for the axial-vector part we find
\begin{equation}
    M_{XV} = \langle V(p_{V})| \bar{Q}_{i}\gamma_{5}Q_{j} |X(p_{X})\rangle = -\frac{(\epsilon^{*}(p_{V}) \cdot p_{X})}{m_{Q_{i}}+m_{Q_{j}}}2m_{V}A^{XV}_{0}(q^{2}),
\end{equation}
where we used the relation~\eqref{eq:a3-a0}. Consider a scalar product $(\epsilon^{*}(p_{V}) \cdot p_{X})$ in the rest frame of the meson $X$. In this case only longitudinal polarization of $\epsilon^{*}_{\mu}(p_{V})$ contributes. Using $\epsilon_{\mu}^{L,*}(p_{V}) = \left(\frac{|\bm{p}_{V}|}{m_{V}},\frac{\bm{p}_{V}}{|\bm{p}_{V}|}\frac{E_{V}}{m_{V}}\right)$ we obtain
\begin{equation}
    M_{XV} = -\frac{2m_{X}|\bm{p}_{V}|}{m_{Q_{i}}+m_{Q_{j}}}A_{0}(q^{2})
\end{equation}

For the case $B\to K^{*}(892)$, we follow~\cite{Ball:2004rg} and parametrize the form-factor as
\begin{equation}
    A^{BK^{*}(892)}_{0}(q^{2}) = \frac{r_{1}}{1-q^{2}/m_{R}^{2}}+\frac{r_{2}}{1-q^{2}/(m_{\text{fit}}^{A_{0}})^{2}}.
    \label{eq:vector-form-factor-1}
\end{equation}
The values of parameters are given in Table~\ref{tab:vector-form-factor-parameters1}.

For the case $B\to V = K^{*}(1410), K^{*}(1680)$, we use an expression for the form-factors~\cite{Lu:2011jm,Hatanaka:2009gb}: 
\begin{equation}
     A^{BV}_{0}(q^{2}) = \left( 1-\frac{2m_{V}^{2}}{m_B^2 + m_V^2 - q^2}\right)\xi_{||}(q^{2})+\frac{m_{V}}{m_{B}}\xi_{\perp}(q^{2}),
     \label{eq:vector-form-factor-2}
\end{equation}
where 
\begin{equation}
\xi_{\perp/||}(q^{2}) = \frac{\xi_{\perp/||}(0)}{1-q^{2}/m_{B}^{2}}
\end{equation}
The values of the parameters are given in  Table~\ref{tab:vector-form-factor-parameters2}.

\begin{table}[t!]
    \centering
    \begin{tabular}{|c|c|c|c|c|c|}
        \hline  
        $V$ & $r_{1}$ & $r_{2}$ & $m_{R}$, GeV & $m_{\text{fit}}, \text{ GeV}$ & $A^{BV}_{0}(0)$ \\
        \hline
        $K^{*}(892)$ & $1.364$ & $-0.99$ & $m_{B}^{+}$ & $\sqrt{36.8}$ & $0.374^{+0.033}_{-0.033}$ \\
        \hline
    \end{tabular}
    \caption{Values of the parameters in the vector form-factor~\eqref{eq:vector-form-factor-1} from~\cite{Ball:2004rg}.}
    \label{tab:vector-form-factor-parameters1}
\end{table}

\begin{table}[t!]
    \centering
    \begin{tabular}{|c|c|c|c|}
        \hline  
        $V$ & $\xi_{\perp}(0)$ & $\xi_{||}(0)$ & $A^{BV}_{0}(0)$ \\
        \hline
        $K^{*}(1410)$ & $0.28^{+0.04}_{-0.04}$ & $0.22^{+0.03}_{-0.03}$ & $0.3^{+0.036}_{-0.036}$ \\
        \hline
        $K^{*}(1680)$ & $0.24^{+0.05}_{-0.05}$  & $0.18^{+0.03}_{-0.03}$ & $0.22^{+0.04}_{-0.04}$ \\
        \hline
    \end{tabular}
    \caption{Values of the parameters in the vector form-factors~\eqref{eq:vector-form-factor-2} from~\cite{Lu:2011jm,Hatanaka:2009gb}.}
    \label{tab:vector-form-factor-parameters2}
\end{table}

\subsubsection{Pseudo-vector}
\label{app:pseudovector}
For the pseudo-vector mesons, $X' = A$, the expansion of the matrix elements is similar to~\eqref{eq:vector-form-factor-axial-part},~\eqref{eq:vector-form-factor-vector-part}, but the expressions for the vector and axial-vector matrix elements are interchanged~\cite{Bashiry:2009wq,Hatanaka:2008gu},
\begin{multline}
    \langle A(p_{A})| \bar{Q}_{i}\gamma^{\mu}Q_{j} |X(p_{X})\rangle = (m_{X}+m_{A})\epsilon^{\mu *}(p_{A})V_{1}(q^{2}) - \\ -(\epsilon^{*}(p_{A})\cdot q)(p_{X} + p_{A})^{\mu}\frac{V_{2}(q^{2})}{m_{X}+m_{A}}-2m_{A}\frac{\epsilon^{*}(p_{A})\cdot q}{q^{2}}q^{\mu}(V_{3}(q^{2})-V_{0}(q^{2})),
    \label{eq:pseudovector-form-factor-axial-part}
\end{multline}
\begin{equation}
     \langle A(p_{A})| \bar{Q}_{i}\gamma^{\mu}\gamma_{5}Q_{j} |X(p_{X})\rangle= \frac{2A(q^{2})}{m_{X}+m_{A}}i\epsilon^{\mu\nu\rho\sigma}\epsilon^{*}_{\nu}(p_{A})p_{X, \rho}p_{A, \sigma},
     \label{eq:pseudovector-form-factor-vector-part}
\end{equation}
with the same relation between $V_i$ as for $A_i$ in the case of vector mesons~\eqref{eq:a3-a0}. We therefore obtain
\begin{equation}
    M_{XA} = \frac{2m_{X}|\bm{p}_{A}|}{m_{Q_{j}}-m_{Q_{i}}}V^{XA}_{0}(q^{2}),
\end{equation}

We will consider two lightest pseudo-vector resonances $K_{1}(1270), K_{1}(1400)$, each of which is the mixture of unphysical $K_{1A}$ and $K_{1B}$ states~\cite{Bashiry:2009wq},
\begin{equation}
    \begin{pmatrix} |K_{1}(1270) \rangle \\ |K_{1}(1400) \rangle \end{pmatrix} =  \begin{pmatrix} \sin(\theta_{K_{1}}) & \cos(\theta_{K_{1}}) \\ \cos(\theta_{K_{1}}) & -\sin(\theta_{K_{1}})\end{pmatrix} 
    \begin{pmatrix} |K_{1A} \rangle \\ |K_{1B}\rangle \end{pmatrix},
\end{equation}
The form-factors $V_{0}^{BK_{1}}$ can be related to the form-factors $V_{0}^{A/B}$ of the $K_{1A}, K_{1B}$ as 
\begin{align}
V^{BK_{1}(1270)}_{0}(q^{2}) =\frac{1}{m_{K_{1}(1270)}}\left[\sin(\theta_{K_{1}})m_{K_{1A}}V_{0}^{A}(q^{2})+\cos(\theta_{K_{1}})m_{K_{1B}}V_{0}^{B}(q^{2})\right], \label{eq:pseudovector-form-factors1} \\
V^{BK_{1}(1400)}_{0}(q^{2}) =\frac{1}{m_{K_{1}(1400)}}\left[\cos(\theta_{K_{1}})m_{K_{1A}}V_{0}^{A}(q^{2})-\sin(\theta_{K_{1}})m_{K_{1B}}V_{0}^{B}(q^{2})\right],
    \label{eq:pseudovector-form-factors2}
\end{align}
where
\begin{equation}
    V_{0}^{A/B}(q^{2}) = \frac{F^{A/B}_{0}}{1-a_{A/B}\frac{q^{2}}{m_{B}^{2}}+b_{A/B}\left(\frac{q^{2}}{m_{B}^{2}}\right)^{2}}.
    \label{eq:pseudovector-form-factors-unphys}
\end{equation}
The values of all relevant parameters are given in Tables~\ref{tab:pseudovector-form-factors-parameters1}, \ref{tab:pseudovector-form-factors-parameters2}. 

\begin{table}[t!]
    \centering
    \begin{tabular}{|c|c|c|c|c|c|}
        \hline  
       $F_{0}^{A}$ & $F_{0}^{B}$ & $a_{A}$ & $a_{B}$ & $b_{A}$ & $b_{B}$ \\
        \hline
        $0.22^{+0.04}_{-0.04}$ & $-0.45^{+0.12}_{-0.08}$ & $2.4$ & $1.34$ & $1.78$ & $0.69$ \\
        \hline
    \end{tabular}
    \caption{Values of the parameters in the vector form-factors~\eqref{eq:pseudovector-form-factors-unphys} from~\cite{Bashiry:2009wq}.}
    \label{tab:pseudovector-form-factors-parameters1}
\end{table}
\begin{table}[t!]
    \centering
    \begin{tabular}{|c|c|c|c|c|}
        \hline  
       $\theta_{K_1}$ & $m_{K_{1A}}$ & $m_{K_{1B}}$ & $V_{0}^{BK_{1}(1270)}(0)$ & $V_{0}^{BK_{1}(1400)}(0)$ \\
        \hline
        $-34^{\circ}\pm 13^{\circ}$ & 1.31\text{ GeV} & 1.34\text{ GeV} & $-0.52^{+0.13}_{-0.09}$ & $-0.07^{+0.033}_{-0.012}$ \\
        \hline
    \end{tabular}
    \caption{Values of the parameters in the vector form-factors~\eqref{eq:pseudovector-form-factors1}, \eqref{eq:pseudovector-form-factors2} from~\cite{Bashiry:2009wq}.}
    \label{tab:pseudovector-form-factors-parameters2}
\end{table}

\subsection{Tensor final meson state}
\label{app:tensor}

For the tensor meson, $X' = T$, the expansion of the matrix element is~\cite{Li:2010ra,Cheng:2010yd}
    \begin{multline}
    \langle T(p_{T})| \bar{Q}_{i}\gamma^{\mu}\gamma_{5}Q_{j} |X(p_{X})\rangle = (m_{X}+m_{T})\epsilon^{\mu *,s}_{T}(p_{T})A_{1}(q^{2}) - \\ -(\epsilon^{*,s}_{T}(p_{T})\cdot q)(p_{X} + p_{T})^{\mu}\frac{A_{2}(q^{2})}{m_{X}+m_{T}}-2m_{T}\frac{\epsilon^{*,s}_{T}(p_{T})\cdot q}{q^{2}}q^{\mu}(A_{3}(q^{2})-A_{0}(q^{2}))
    \label{eq:tensor-form-factor-axial-part}
\end{multline}
Here, $\epsilon^{s}_{T\mu}(p_{T})$ is a vector defined by
\begin{equation}
    \epsilon_{T\mu}^{s}(p_{T}) \equiv  \frac{1}{m_{X}}\epsilon_{\mu\nu}^{s}(p_{T})p_{X}^{\nu},
\end{equation}
with $\epsilon_{\mu\nu}^{s}$ being the polarization tensor of $T$ satisfying $p_{\mu}\epsilon^{\mu\nu,s}(p) = 0$ and $\epsilon^{\mu\nu,s} = \epsilon^{\nu\mu,s}$, $\epsilon^{\mu, \ s}_{\ \mu} = 0$. For particular polarizations $s = \pm 2, \pm 1, 0$ we have~\cite{Li:2010ra}
\begin{equation}
    \epsilon^{\pm 2}_{T\mu} = 0, \quad \epsilon_{T\mu}^{\pm 1} = \frac{1}{m_{h}\sqrt{2}}(\epsilon^{0}\cdot p_{X})\epsilon^{\pm 1}_{\mu}, \quad \epsilon^{0}_{T\mu} = \sqrt{\frac{2}{3}}\frac{\epsilon^{0}\cdot p_{X}}{m_{X}}\epsilon_{\mu}^{0},
\end{equation}
where
\begin{equation}
    \epsilon^{\pm 1}_{\mu} = \frac{1}{\sqrt{2}}(0, \mp 1, i, 0), \quad \epsilon_{\mu}^{0} = \frac{1}{m_{T}}(|\bm{p}_{T}|,0,0,E_{T}).
\end{equation}
Repeating the same procedure as in the previous section, we find that to $q_{\mu}M^{\mu,s}_{XT}$ contributes only the polarization $s = 0$, and therefore
\begin{equation}
    M_{XT}  = -\frac{q_{\mu}M^{\mu, 0}_{XT}}{m_{Q_{i}}+m_{Q_{j}}}= -\frac{1}{m_{Q_{i}}+m_{Q_{j}}}\sqrt{\frac{2}{3}}\frac{m_{X}|\bm{p}_{T}|^{2}}{m_{T}}2 A^{XT}_{0}(q^{2}).
\end{equation}
The parametrization of the form-factor $A^{XT}_{0}$ is~\cite{Li:2010ra,Cheng:2010yd}
\begin{equation}
A^{XT}_{0}(q^{2}) = \frac{F^{XT}_{0}}{\left( 1-\frac{q^{2}}{m_{X}^{2}}\right)\left( 1-a_{T}\frac{q^{2}}{m_{X}^{2}}+b_{T}\left(\frac{q^{2}}{m_{X}^{2}}\right)^{2}\right)}
\end{equation}
For the transition $B\to K_{2}^{*}(1430)$ we use the values $F^{BK_{2}^{*}}_{0} = 0.23$, $a_{T} = 1.23$, $b_{T} = 0.76$ from~\cite{Cheng:2010yd}.

\section{Production from mesons through quartic coupling}
\label{sec:quartic-coupling}
The quartic coupling 
\begin{equation}
    \mathcal{L}_{\text{quartic}} = \frac{\alpha}{2}hS^{2}
    \label{eq:quartic-coupling}
\end{equation}
generates new production channels from the mesons
\begin{equation}
X_{Q_i} \to X'_{Q_j}SS, \quad X\to SS,
\label{eq:quartic-coupling-processes}
\end{equation}
that are described by Feynman diagrams in Fig.~\ref{fig:Sproduction-quartic} (b). 

The matrix element for decays $X_{Q_{i}}\to X'_{Q_{j}}SS$ can be written in terms of the matrix element $M_{XX'}$ of hadronic transitions given by Eq.~\eqref{eq:hadronic-matrix-element-1}:
\begin{equation}
    \mathcal{M}(X_{Q_{i}} \to X'_{Q_{j}}SS) \approx \frac{\alpha}{m_{h}^{2}}\frac{m_{Q_i}}{2v}\xi_{ij}M_{XX'}(q^{2}),
\end{equation}
where $q^{2}$ is invariant mass of scalars pair, $M_{XX'}(q^{2})$ is the matrix element of hadronic transitions $X_{Q_{i}} \to X'_{Q_{j}}$ given by Eq.~\eqref{eq:hadronic-matrix-element-1}.

The matrix element for a process $X_{Q_{i}Q_{j}} \to SS$ can be expressed in terms of the decay constant $f_{X}$ of the meson $X$. Namely, $f_{X}$ is defined by
\begin{equation}
    \langle 0|\bar{Q}_{i}\gamma_{\mu}\gamma_{5}Q_{j} |X(p)\rangle \equiv if_{X}p_{\mu}
\end{equation}
Contracting it with $p_{\mu}$ and using the same trick as in Eq.~\eqref{eq:trick}, we obtain
\begin{equation}
    \langle 0|\bar{Q}_{i}\gamma_{5}Q_{j} |X(p)\rangle \equiv -\frac{if_{X}m_{X}^{2}}{m_{Q_{i}} - m_{Q_{j}}}
\end{equation}
Therefore, the matrix element $\mathcal{M}(X_{Q_{i}Q_{j}} \to SS)$ is
\begin{equation}
    \mathcal{M}(X_{Q_{i}Q_{j}} \to SS)= \frac{m_{Q_{i}}\xi_{ij}}{2v m_{h}^{2}} \langle 0|\bar{Q}_{i}\gamma_{5}Q_{j} |X(p)\rangle \approx i\frac{\alpha f_{X}m_{X}^{2}}{2v m_{h}^{2}}\xi_{ij},
\end{equation}
The values of $f_{X}$ are summarized in Table~\ref{tab:meson-decay-constants}.
\begin{table}[]
    \centering
    \begin{tabular}{|c|c|c|c|c|c|}
    \hline    
    Meson $X$ & $B_{0}$ & $B_{s}$ & $K_{0}$ \\
    \hline
    $f_{X},\text{ GeV}$ & $0.19$ & $0.23$ & $0.16$ \\
    \hline
    \end{tabular}
    \caption{Values of meson decay constants. We use~\cite{Chang:2018aut} and references therein.}
    \label{tab:meson-decay-constants}
\end{table}
For the decay width of the process $X_{Q_{i}Q_{j}} \to SS$ we find
\begin{equation}
    \Gamma (X_{Q_{i}Q_{j}}\to SS) = \frac{m_{X}^3}{v^2} \frac{|\xi_{ij}|^2 f_{X}^2 \alpha^2}{128\pi m_h^4}\sqrt{1-\frac{4m_{S}^{2}}{m_{X}^{2}}}
    \label{eq:bstossbranching}
\end{equation}
The decay width for the process $X_{Q_{i}} \to X'_{Q_{j}}SS$ can be calculated using the formulas from Appendices~\ref{sec:inclusive-br-calculation}. Namely, we have
\begin{equation}
   \Gamma_{X_{Q_{i}} \to X'_{Q_{j}}SS} = \frac{|\xi_{ij}|^{2}m_{Q_{i}}^{2}\alpha^{2}}{512\pi^{3}m_{X}^{3}v^{2}m_{h}^{4}}\int \limits_{4m_{S}^{2}}^{(m_{X}-m_{X'})^{2}} |M_{XX'}(q^{2})|^{2}\sqrt{(E_{2}^{*})^{2}-m_{S}^{2}}\sqrt{(E_{3}^{*})^{2}-m_{X'}^{2}}dq^{2},
   \label{eq:br-quartic-3-body}
\end{equation}
where $q^{2}$ is the squared invariant mass of two scalars, and
\begin{equation}
   E_{2}^{*} =\frac{\sqrt{q^{2}}}{2}, \quad E_{3}^{*} = \frac{m_{X}^{2}-q^{2}-m_{X'}^{2}}{2\sqrt{q^{2}}}
\end{equation}

\section{Decays of a scalar}
\label{sec:decay}

\subsection{Decay into leptons and photons}
\label{sec:decayleptonphoton}

The decay width of the $S$ particle into leptons pair simply follows from the Lagrangian~\eqref{eq:L1} and reads
\begin{equation}
\Gamma (S \rightarrow l^+ l^-) 
= \frac{ \theta^2 y_{f}^{2} m_S}{8 \pi} \beta_{l}^{3},
\label{eq:Gll}
\end{equation}
where $\beta_{l} = \left( 1 - \frac{4 m_l^2}{m_S^2} \right)^{1/2}$.
The decay width into photons is
\begin{equation}
    \Gamma (S \rightarrow \gamma \gamma) =
    |F_{\gamma}(m_{S})|^2 \left( \frac{\alpha_{\text{EM}}}{8 \pi} \right)^2 \frac{\theta^2 m_S^3}{8 \pi v^2},
    \label{eq:Sto2gamma}
\end{equation}
Where $F_{\gamma}$ is given by Eq.~\eqref{eq:photon-gluon-loop-factor}.

\subsection{Decays into quarks and gluons}
\label{sec:decay-perturbative}
The decay width into quarks in leading order in $\alpha_{s}$ can be obtained directly from the Lagrangian~\eqref{eq:L1}; the QCD corrections were obtained in~\cite{Spira:1997dg}. In order to take into account the quark hadronization, we follow~\cite{Winkler:2018qyg,Alekhin:2015byh,Gunion:1989we} and use the mass of the lightest hadron $m_{M_{q}}$ containing quark $q$ instead of the quark mass $m_{q}$ in the kinematical factors. The result is
\begin{equation}
    \Gamma (S \rightarrow \bar{q} q) 
    = N_c \frac{  \theta^2 m_S \overline{m}_q^2(m_S)}{8 \pi v^2} \left( 1 - \dfrac{4 \overline{m}_{M_{q}}^2}{m_S^2} \right)^{3/2} 
    \big(1 + \Delta_{\text{QCD}} + \Delta_t\big),
    \label{eq:decay-quarks}
\end{equation}
where $M_{q} = K$ for the $s$ quark and $D$ for $c$ quark, the factor $N_c = 3$ stays for the number of the QCD colors,\vspace{-1em}
\begin{align}
    \Delta_{\text{QCD}} &= 
    5.67 \frac{\alpha_s(m_S)}{\pi} + (35.94 - 1.36 N_f) \left(\frac{\alpha_s(m_S)}{\pi}\right)^2 
    \nonumber\\
    &+ (164.14 - 25.77 N_f + 0.259 N_f^2) \left(\frac{\alpha_s(m_S)}{\pi}\right)^3, \\
    \Delta_t &= \left(\frac{\alpha_s(m_S)}{\pi}\right)^2 
    \left( 1.57 - \frac{2}{3}
    \text{log}\frac{m_S^2}{m_t^2} + \frac{1}{9} \text{log}^2 \frac{\overline{m}_q^2(m_S)}{m_S^2}\right),
\end{align}
and the running mass \cite{Spira:1997dg} $\overline{m}_q(m_S)$ is given by\vspace{-0.5em}
\begin{equation}
    \overline{m}_q(m_S) = 
    \overline{m}_q(Q) 
    \frac{c(\alpha_s(m_S)/\pi)}{c(\alpha_s(Q)/\pi)},
\end{equation}
with the coefficient $c$, which is equal to\vspace{-0.5em}
\begin{align}
    c(x) &= \left(\frac{9}{2} x\right)^{4/9} 
    (1 + 0.895 x + 1.371 x^2 + 1.952 x^3), 
    &\text{ for }
    &m_s < m_S < m_{c}, \\
    c(x) &= \left(\frac{25}{6} x\right)^{12/25} (1 + 1.014 x + 1.389 x^2+ 1.091 x^3), 
    &\text{ for }
    &m_{c} < m_S < m_{b},\\
    c(x) &= \left(\frac{23}{6} x\right)^{12/23} (1 + 1.175 x + 1.501 x^2 + 0.1725 x^3), &\text{ for }
    &m_{b} < m_S < m_t.
\end{align}
We use the MS-mass at $Q = 2 \text{ GeV}$ scale \cite{Sanfilippo:2015era}: $\overline{m}_c = 1.23 \text{ GeV}$ and $\overline{m}_s = 0.0924 \text{ GeV}$.

For decays into gluons, using the effective couplings~\eqref{eq:dis-lagrangian}, summing over all gluon species (which gives a factor of 8) and including QCD corrections, we obtain~\cite{Spira:1997dg}
\begin{equation}
    \Gamma (S \rightarrow GG) = 
    |F_{G}(m_{S})|^2 \left( \frac{\alpha_s}{4 \pi} \right)^2 \frac{\theta^2 m_S^3}{8 \pi v^2} \left( 1 + \frac{m_t^2}{8 v^2 \pi^2} \right),
    \label{eq:decay-gluons}
\end{equation}
Where $F_{G}$ is given by Eq.~\eqref{eq:photon-gluon-loop-factor}.

\bibliographystyle{JHEP}
\bibliography{ship.bib}

\end{document}